\documentclass[10pt,aps,reprint,amsmath,amssymb,floatfix,citeautoscript,showpacs,twocolumn,notitlepage,prl]{revtex4-1}
\usepackage{mathtools}
\usepackage{txfonts}

\usepackage[hidelinks]{hyperref}
\usepackage{dsfont}
\usepackage{graphicx}     
\usepackage{tikz}
\usepackage[range-phrase=--,range-units=single]{siunitx}

\usepackage{bm}
\usepackage{standalone}
\usepackage{xr}

\tikzset{>=stealth}

\newcommand{\citeprefix}{}
\newcommand{\mycite}[1]{\cite{\citeprefix#1}}
\newcommand{\mybibliography}[1]{\bibliography{#1}}

\newcommand{\bs}[1]{{\bm{#1}}} 
\newcommand{\E}{\mathds{1}}
\renewcommand\d{\mathop{}\!\mathrm{d}}
\renewcommand{\i}{\mathrm{i}}

\newcommand{\zz}{\mathbb{Z}_2}
\newcommand{\bk}{\bs{k}}

\newcommand{\ef}{\epsilon_{f}}
\newcommand{\cn}{C_0}
\newcommand{\cp}{C_\pi}
\newcommand{\cd}{C_\mathrm{d}}
\newcommand{\Eg}{$\mathrm E_g$}
\newcommand{\smb}{SmB$_6$}

\newcommand{\fig}[1]{\autoref{#1}}
\newcommand{\subfig}[2]{\hyperref[#1:#2]{\autoref*{#1}\,\ref*{#1:#2}}}
\newcommand{\subfigg}[2]{\hyperref[#1:#2]{\ref*{#1}\,\ref*{#1:#2}}}
\newcommand{\subfigs}[3]{Figs.~\subfigg{#1}{#2} and \subfigg{#1}{#3}}

\newcommand{\eq}[1]{\hyperref[#1]{Eq.~(\ref*{#1})}}
\newcommand{\Eq}[1]{\hyperref[#1]{Equation~(\ref*{#1})}}
\newcommand{\tab}[1]{\autoref{#1}}
\makeatletter
\newcommand{\Ref}[1]{%
\protect\hyper@link{cite}{cite.\citeprefix #1}%
{Ref.~\protect\NoHyper\onlinecite{\citeprefix #1}\protect\endNoHyper}}
\makeatother
\renewcommand{\eqref}[1]{\hyperref[#1]{(\ref*{#1})}}
\newcommand{\page}[1]{\hyperref[#1]{page~\pageref*{#1}}}

\DeclareMathOperator{\sign}{sgn}
\newcommand{\updown}{\uparrow\downarrow}

\newcommand{\abb}[1]{{\MakeUppercase{\small #1}}}
\newcommand{\tki}{\abb{tki}}
\newcommand{\arpes}{\abb{arpes}}

\newcommand{\hsp}{\abb{hsp}}
\newcommand{\hsl}{\abb{hsl}}
\newcommand{\mil}{\abb{mil}}
\newcommand{\bz}{\abb{bz}}
\newcommand{\sbz}{s\abb{bz}}
\newcommand{\nn}{\abb{nn}}
\newcommand{\nnn}{\abb{nnn}}
\newcommand{\mcn}{\abb{mcn}}
\newcommand{\sm}{Supplemental Material~\mycite{sm}}

\newcommand{\pt}[1]{\mathrm{#1}}
\newcommand{\ptt}[1]{$\pt{#1}$}
\newcommand{\g}{\ptt\Gamma}
\newcommand{\x}{\ptt X}
\newcommand{\m}{\ptt M}
\renewcommand{\r}{\ptt R}

\newcommand{\dxsys}{$(x^2-y^2)$}
\newcommand{\dzsrs}{$(3z^2-r^2)$}

\newcommand{\ket}[1]{\left|#1\right\rangle}
\newcommand{\transp}{^\mathrm{t}}

\let\Im\relax
\DeclareMathOperator{\Im}{Im}
\renewcommand{\mod}{\;\text{mod}\;}

\newcommand{\onlyonecol}[1]{}
\newcommand{\onlytwocol}[1]{}
\newcommand{\onetwocol}[2]{\onlyonecol{#1}\onlytwocol{#2}}

\newcommand{\mysection}[1]{\smallskip\emph{#1 ---}\phantomsection}

\makeatletter
\newcommand{\customlabel}[2]{%
   \protected@write \@auxout {}{\string \newlabel {#1}{{#2}{\thepage}{#2}{#1}{}} }%
   \hypertarget{#1}{#2}
}
\makeatother

\newcommand{\thisfig}{}
\newsavebox{\temptikzbox}
\newcommand{\figlabel}[3][0,0]{%
	\sbox{\temptikzbox}{#2}
	\begin{tikzpicture}[inner sep=0pt]
		\node [anchor=north west] at (0,0) {\usebox\temptikzbox};
		\node [anchor=north west] at (#1) {\customlabel{\thisfig:#3}{(#3)}};
	\end{tikzpicture}
}

\graphicspath{{Figures/}}
\newcommand{\mytitle}{Surface-state spin textures and mirror Chern numbers in topological Kondo insulators}
\newcommand{\affeth}{%
	\affiliation{%
		Institut f\"ur Theoretische Physik, ETH Z\"urich,
		8093 Z\"urich, Switzerland
	}
}
\hypersetup{
	pdfauthor={Markus Legner, Andreas R\"uegg, Manfred Sigrist},
	pdftitle={\mytitle},
	pdfcreator={Markus Legner}
}

\newcounter{mypart}
\makeatletter
\@addtoreset{equation}{mypart}
\@addtoreset{figure}{mypart}
\@addtoreset{table}{mypart}
\@addtoreset{page}{mypart}
\makeatother

\begin{document}
\stepcounter{mypart}
\setcounter{page}{1}
\title{\mytitle}

\author{Markus Legner} 
\affeth
\author{Andreas R\"uegg}
\affeth
\author{Manfred Sigrist} 
\affeth
\date{\today}
% \doi{10.1103/PhysRevLett.115.156404}

%-----------------------------------------------------%
%-----------------------------------------------------%
\begin{abstract}
	The recent discovery of topological Kondo insulators has triggered renewed interest in the well-known Kondo insulator samarium hexaboride, which is hypothesized to belong to this family.
	In this Letter, we study the spin texture of the topologically protected surface states in such a topological Kondo insulator.
	In particular, we derive close relationships between (i) the form of the hybridization matrix at certain high-symmetry points, (ii) the mirror Chern numbers of the system, and (iii) the observable spin texture of the topological surface states. In this way, a robust classification of topological Kondo insulators and their surface-state spin texture is achieved.
	We underpin our findings with numerical calculations of several simplified and realistic models for systems like samarium hexaboride.
\end{abstract}

\maketitle

%-----------------------------------------------------%
%-----------------------------------------------------%
\mysection{Introduction}\label{s:intro}
Since the theoretical characterization of topological Kondo insulators (\tki s)~\mycite{dzero_topological_2010,dzero_theory_2012}, this class of materials has attracted much attention in the community. 
One material in particular, samarium hexaboride (\smb), has been studied extensively both theoretically and experimentally.
Several theoretical studies predicted \smb~\mycite{takimoto_smb6_2011,tran_phase_2012,lu_correlated_2013,alexandrov_cubic_2013,dzero_new_2013,ye_tci_2013} and related compounds~\mycite{deng_plutonium_2013,weng_correlated_2013} to be \tki{}s with protected gapless surface modes.
Different experiments showed that, at sufficiently small temperatures, transport is indeed dominated by the surface contributions~\mycite{wolgast_low_2013,kim_robust_2013,zhang_hybridization_2013}. 
At the same time, angle-resolved-photoemission-spectroscopy (\arpes)~\mycite{miyazaki_momentum_2012,xu_surface_2013,neupane_surface_2013,jiang_observation_2013,frantzeskakis_kondo_2013,min_importance_2013}, quantum-oscillation~\mycite{li_quantum_2013}, and scanning-tunneling-microscopy measurements~\mycite{yee_imaging_2013} confirmed the existence of gapless surface states. 
Nevertheless, due to the small bulk gap of \SIrange{15}{20}{\milli\eV}~\mycite{gorshunov_smb6gap_1999,xu_spin_2014,hlawenka_trivial_2015} and strong electronic correlations, a detailed characterization of the nature of the surface states is difficult and may require additional concepts such as atomic reconstruction~\mycite{heming_surface_2014}, Kondo breakdown~\mycite{alexandrov_breakdown_2015}, or excitonic scattering~\mycite{kapilevich_incomplete_2015}. 
Some groups also challenged the scenario of a \tki~\mycite{zhu_polarity_2013,hlawenka_trivial_2015}.
To date, the most conclusive evidence for the topological nature of the surface states is provided by spin-resolved \arpes{} measurements of the (001) surface~\mycite{xu_spin_2014} showing that the surface states around the \ptt{\bar X} point of the surface Brillouin zone (\sbz) are spin-polarized. 

\begin{figure}[tb]
	\renewcommand{\thisfig}{fig:spin-textures}
	\centering
		
	\figlabel{\includegraphics{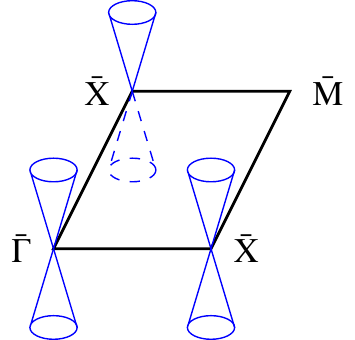}}{a}\hfill
	\figlabel{\includegraphics{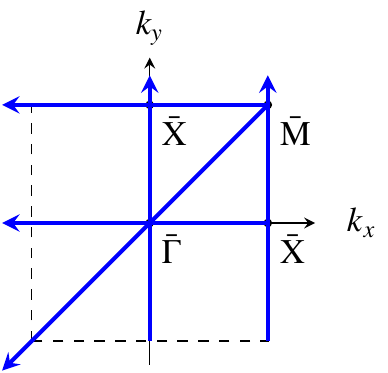}}{b}\\[5mm]
	
	\figlabel{\includegraphics{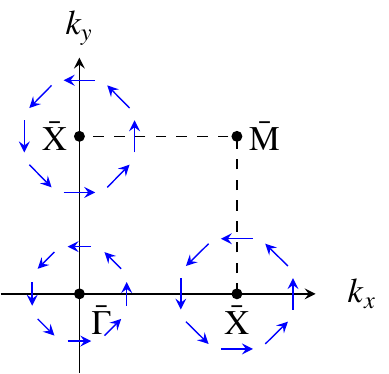}}{c}\hfill
	\figlabel{\includegraphics{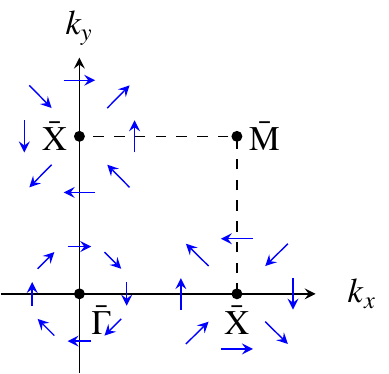}}{d}
	\caption{{(a) Sketch of the Dirac cones in the (001) surface Brillouin zone. 
	(b) Positive directions of mirror invariant lines in the \sbz{} for the (001) surface with outward pointing normal vector $\bs n_\mathrm{sf}=\bs e_z$, see {\Ref{sm}} for further details. 
	(c-d) Sketch of the spin (or pseudospin, see \page{s:pseudospin}) textures in the (001) \sbz.
	While at \ptt{\bar\Gamma} the winding number is always $w_{\pt{\bar\Gamma}}=1$, at the \ptt{\bar X} points it can be $w_{\pt{\bar X}}=+1$ (c) or $w_{\pt{\bar X}}=-1$ (d), depending on the configuration of the \mcn{}s.}}
	\label{\thisfig}
\end{figure}

\smb{} is predicted to have a band inversion at the \ptt{X} high-symmetry points (\hsp s)~\mycite{takimoto_smb6_2011,tran_phase_2012,lu_correlated_2013,alexandrov_cubic_2013}.
The \ptt{X}-inverted phase has a nontrivial strong $\zz$ index $\nu_0=1$, weak topological indices $\bs\nu=(1,1,1)$ and protected surface Dirac cones as shown in \subfig{fig:spin-textures}{a} for the (001) surface. 
The experimental work in \Ref{xu_spin_2014} is consistent with these predictions and furthermore suggests that the spin texture of the surface states is as sketched in \subfig{fig:spin-textures}{c}.
Interestingly, however, several theoretical studies reached conflicting conclusions about the nature of the spin texture~\mycite{yu_model_2014, kim_termination_2014, baruselli_scanning_2014}, which is not uniquely determined by the $\zz$ invariants.
In fact, for linear Dirac cones, two situations are compatible with the cubic symmetry, see \subfigs{fig:spin-textures}{c}{d}.
They are distinguished by opposite winding numbers $w_{\pt{\bar X}}=\pm 1$ of the planar unit spin $(n_x,n_y)=(S_x,S_y)/\sqrt{S_x^2+S_y^2}$ around the \ptt{\bar X} point, where the winding number around the \hsp{} ${\bs K}$~\footnote{The \hsp{} $\bs K$ has the property $-\bs K=\bs K+\bs G$, where $\bs G$ is a reciprocal lattice vector. On the (001) surface, there are three different \hsp{}s: \ptt{\bar\Gamma}, \ptt{\bar X}, and \ptt{\bar M}.} is defined as
\begin{equation}
w_{\bs K}=\frac{1}{2\pi}\oint_{\gamma_{\bs K}} \bs{\nabla}\left[\Im\log(n_x+\i n_y)\right]\cdot \d\bs{s}\,,
\end{equation}
with $\gamma_{\bs K}$ a contour encircling ${\bs K}$ in an anticlockwise fashion.
This discrepancy between different theoretical models and approaches raises the important question of what determines the spin texture in cubic \tki{}s.

In this Letter, we provide two answers to this question: First, we show that there is a close connection between the spin texture and the mirror Chern numbers (\mcn s)~\mycite{fu_topological_2011}. 
In particular, knowledge of the \mcn{}s allows us to distinguish between the two situations shown in \subfigs{fig:spin-textures}{c}{d}.
Second, we provide analytical expressions relating the surface-state spin texture to the hybridization parameters of specific models. 
These relations demonstrate that the number and type of included orbitals in the effective model does \emph{not} uniquely define the winding number; instead, the relative strength of different-range hybridization parameters is equally important. 
In addition, we show how the system can be tuned across topological phase transitions, during which the surface-state spin texture changes while all the $\zz$ invariants remain unaffected. 

In the remainder of the Letter, we will provide the details to the above statements.
We will also apply the general argumentation to a multiorbital model with itinerant \Eg{} and localized $\Gamma_8$ electrons, as in \Ref{baruselli_scanning_2014}.
Other models are discussed in the \sm{}.

\mysection{Mirror Chern numbers define pseudospin texture}\label{s:pseudospin}
To start, we review certain facts about the \mcn{}s in \smb{}.
The \mcn{}s are topological invariants, which are protected by mirror symmetries~\mycite{fu_topological_2011,ye_tci_2013,legner_topological_2014}. 
In a cubic system, there are three distinct \mcn s: 
$\cn \equiv C^{(+\i)}_{k_\alpha=0}$, $\cp \equiv C^{(+\i)}_{k_\alpha=\pi}$, and $\cd \equiv C^{(+\i)}_{k_\alpha=k_\beta}$, with $\alpha,\beta\in\{x,y,z\}$ and $\beta\neq\alpha$, where $C^{(+\i)}_{\mathcal{S}}$ refers to the Chern number of the Bloch states on the mirror-invariant plane $\mathcal{S}$ with eigenvalue $+\i$ under the mirror operation, see also \Ref{legner_topological_2014}. 
As was shown in \Ref{ye_tci_2013}, the cubic symmetry implies that the \mcn{}s in the \ptt{X}-inverted phase are $\cn =2 \mod 4$, $\cp  = 1 \mod 4$ and $\cd  = 1 \mod 2$. 
These values imply two additional Dirac nodes along the \ptt{\bar{\Gamma}\bar{X}} line on the (110) surface~\mycite{ye_tci_2013,legner_topological_2014}. In the following, we show that the \mcn{}s also determine the spin texture on the (001) surface. (A related argument for Hg-based topological insulators was presented in \Ref{wang_spin_2014}.)

\begin{figure}[t!]
	\renewcommand{\thisfig}{fig:mcn}
	\centering
	
	\figlabel[1.8,0.2]{\includegraphics{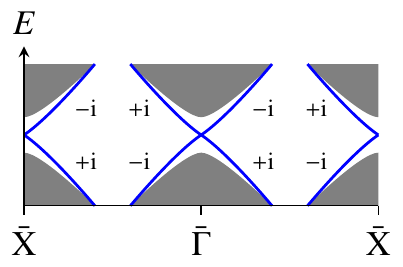}}{a} \hfill
	\figlabel[1.8,0.2]{\includegraphics{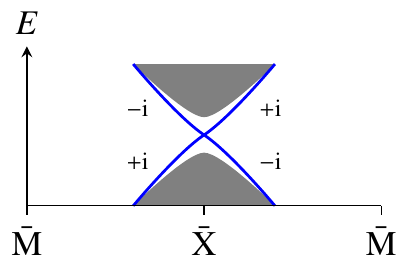}}{b}
	
	\caption{Chiral surface states with mirror eigenvalues $\pm\i$ along the \ptt{\bar\Gamma\bar X} (a) and \ptt{\bar X\bar M} line (b) in positive direction (see \fig{fig:spin-textures}) for $\cn=-2$ and $\cp=1$.}
	\label{\thisfig}
\end{figure}

The projections of the mirror planes onto the (001) surface correspond to the high-symmetry lines (\hsl{}s) shown in \subfig{fig:spin-textures}{b}. 
Along these mirror invariant lines (\mil s), we can classify the surface states according to their mirror-eigenvalues $\pm\i$. 
The bulk-edge correspondence for each mirror-invariant plane then states that the \mcn{} $C$ is equal to the number of right-moving ($C>0$) or left-moving ($C<0$) surface modes with mirror-eigenvalue $+\i$, see \fig{fig:mcn}. 
There exists a certain freedom to choose signs in the calculation of the \mcn{}s. We use a convention~\mycite{sm}, which leads to the positive directions shown in \subfig{fig:spin-textures}{b}. 

The mirror eigenvalues also define a \emph{pseudospin  of the surface states} $\bs\mu$ in the following way: 
On the $k_y=0$ or $k_y=\pi$ \mil, we choose a basis $\{\bs u_{1},\bs u_2\}$ in which the mirror operator takes the form $M_y=-\i\mu_y$, where $\mu_\alpha$ is the $\alpha$-th Pauli matrix. 
Furthermore, on the $k_x=0$ and $k_x=\pi$ \mil{}s we can choose the mirror operator $M_x=-\i\mu_x$.
The pseudospin is then given by the spinor $\bs u = a\bs u_1 + b\bs u_2\equiv (a,b)\transp$. Its relation to the physical spin of the electron is detailed on \page{s:spin}.
It follows that, along the \mil{}s, the pseudospin lies in the surface plane and is always perpendicular to the \mil{}. 
In order to make the connection to the pseudospin texture, it is useful to consider the effective Hamiltonian close to the Dirac node at the \hsp{} ${\bs K}=\pt{\bar{\Gamma}}$ or ${\bs K}=\pt{\bar{X}}$:
\begin{equation}
H_{\bs K}({\bs q})=v_{\bs K}^x\mu_y q_x-v_{\bs K}^y\mu_x q_y= \i(v_{\bs K}^xM_y q_x-v_{\bs K}^yM_x q_y)\,.
\end{equation}
Here, we measure the momentum relative to the respective \hsp, ${\bs q}={\bs k}-{\bs K}$.
At the \ptt{\bar\Gamma}-point, the cubic symmetry implies that $v_{\pt{\bar{\Gamma}}}^x=v_{\pt{\bar{\Gamma}}}^y$ and the resulting pseudospin texture necessarily has a winding number $w_{\bar\Gamma}=1$. 
But at the \ptt{\bar{X}} points, $v_{\pt{\bar X}}^x\neq v_{\pt{\bar X}}^y$ in general, and the winding number of the pseudospin texture is $w_{\pt{\bar X}}=\sign(v_{\pt{\bar X}}^xv_{\pt{\bar X}}^y)$. 
Because the \mcn{}s fix the direction of the pseudospin at the points where the Fermi lines cross the \mil{}s, the \mcn{}s also fix the relative sign between $v_{\pt{\bar X}}^x$ and $v_{\pt{\bar X}}^y$ and hence the winding number $w_{\pt{\bar X}}$. 
It is then easy to see that the set $(\cn,\cp)= (2,1)$ implies the pseudospin texture shown in \subfig{fig:spin-textures}{c}, while $(\cn,\cp)= (-2,1)$ implies the pseudospin texture shown in \subfig{fig:spin-textures}{d}. 
For linear Dirac cones at \ptt{ \bar\Gamma} and \ptt{\bar X}, there are no other possibilities, i.e.~higher \mcn{}s imply additional Dirac nodes along \hsl{}s~\mycite{sm}. 
(Note that in \fig{fig:spin-textures} we assume a chemical potential above the Dirac nodes.)

\mysection{Hybridization matrix defines mirror Chern numbers}\label{s:mcn}
We now analyze the connection between microscopic parameters of the electronic Hamiltonian and the set of \mcn{}s.
From \emph{ab-initio} calculations~\mycite{lu_correlated_2013,kang_band_2013,yu_model_2014,baruselli_scanning_2014} it is known that the states near the Fermi energy  in \smb{} are predominantly formed by the Sm $5d$ electrons of \Eg{} symmetry and the Sm $4f$ electrons in the $J=5/2$ multiplet.
The latter splits further into a $\Gamma_8$ quartet, $\ket{\Gamma_{8,\pm}^{(1)}}=\sqrt{\frac56}\ket{\pm\frac52}+\sqrt{\frac16}\ket{\mp\frac32}$ and $\ket{\Gamma_{8,\pm}^{(2)}}=\ket{\pm\frac12}$, and a $\Gamma_7$ doublet, $\ket{\Gamma_{7,\pm}}=\sqrt{\frac16}\ket{\pm\frac52}-\sqrt{\frac56}\ket{\mp\frac32}$, where the index $\pm$ is the {\em orbital pseudospin}~\footnote{This orbital pseudospin should not be confused with the surface-state pseudospin defined \hyperref[s:pseudospin]{before}. The detailed relation between the orbital pseudospin and the physical spin of the $f$ electrons is given in \Ref{sm}.}.
Our strategy is to start in the trivial insulating phase without band inversion and consider the effective model, which describes the gap closing and subsequent band inversion at the \ptt{X} points.

The little co-group at the \ptt{X} point is isomorphic to the tetragonal symmetry group $\mathrm{D_{4h}}$.
Thus, all the irreducible representations are at most two-dimensional and the band inversion occurs between the energetically highest single Kramers pair of $f$~electrons $f_{\pt{X},\pm}$ and the energetically lowest single Kramers pair of $d$~electrons $d_{\pt{X},\updown}$. 
Near the transition between the trivial and the topological phase, the low-energy electronic structure can be obtained from an effective $4\times4$ Bloch Hamiltonian around the \ptt{X} points,
\begin{equation}
{H}_{\mathrm{eff}}^\pt{X}({\bs q})=\begin{pmatrix}
{\varepsilon}_{\bs q}^d& {\Phi}^{\dag}_{\bs q}\\
{\Phi}^{\vphantom\dag}_{\bs q} & {\varepsilon}_{\bs q}^f
\end{pmatrix}\,.
\label{eq:HeffX}
\end{equation}
\Eq{eq:HeffX} is given for a spinor $\psi=(d_{{\pt X},\uparrow},d_{{\pt X},\downarrow},f_{{\pt X},+},f_{{\pt X},-})\transp$ and ${\bs q}$ is measured from \ptt{X}. 
The simultaneous presence of inversion and time-reversal symmetry allows us to choose the hybridization matrix in the form ${\Phi}_{\bs q}=\i\bs{\phi}_{\bs q}\cdot\bs{\sigma}$, with $\bs{\phi}_{\bs q}=\bs{\phi}^{*}_{\bs q}=-\bs{\phi}_{-\bs q}$ and $\bs \sigma$ the Pauli matrices in spin space. 
In the following, we consider $\pt{X}=(0,0,\pi)$, and expand to lowest order in ${\bs q}$: ${\varepsilon}_{\bs q}^d=\varepsilon_d\E$, ${\varepsilon}_{\bs q}^f=\varepsilon_f\E$ and 
\begin{equation}
{\Phi}_{\bs q}=\i\left[\phi_1(\sigma_xq_x+\sigma_yq_y)+\phi_2\sigma_zq_z\right]\,.
\label{eq:Phi_lin}
\end{equation}
As we show below, the relative sign between the two independent parameters $\phi_1$ and $\phi_2$ of the linearized hybridization matrix~\eqref{eq:Phi_lin} 
determines the set of \mcn{}s and hence the surface-state spin texture in the \ptt{X}-inverted phase. 

First, we address the \mcn{}  $\cn $ and therefore consider the mirror plane $k_x=0$.
The mirror operator in the basis of \eq{eq:HeffX} is $M_x=-\i\tau_z\otimes\sigma_x$. 
Thus, in the subspace $M_x=+\i$, \eq{eq:HeffX} reduces to
\begin{equation}
H^{(+\i)}_{\mathrm{eff},k_x=0}({\bs q})=\bar{\varepsilon}\E-\phi_1q_y\mu_x+\phi_2q_z\mu_y-\Delta\,\mu_z \,,
\end{equation}
where $\mu_\alpha$ are the Pauli matrices acting on the basis vectors $(1,-1,0,0)/\sqrt{2}$ and $(0,0,1,1)/\sqrt{2}$
and we have defined $\bar{\varepsilon}\equiv \frac12(\varepsilon_d+\varepsilon_f)$ and $\Delta\equiv \frac12(\varepsilon_f-\varepsilon_d)$.
The total Berry flux contribution of the lower band of a Dirac model $h(\bs k)=\epsilon(\bs k)\,\hat{\bs d}(\bs k)\cdot\bs \mu$ with $\hat{\bs d}=\bs d/|\bs d|$ is 
\begin{equation}
	C^\mathrm{Dirac}=\frac{1}{4\pi}\int \d k_1 \d k_2\; \hat{\bs{d}}(\bs k)\cdot \left(\frac{\partial \hat{\bs d}}{\partial k_1}\times \frac{\partial \hat{\bs d}}{\partial k_2}\right)\,,
\end{equation}
which for our case with $\bs d(\bs k)=(-\phi_1k_1,\phi_2 k_2,-\Delta)$ leads to 
\begin{equation}
C_{k_x=0}^\mathrm{Dirac}=\frac12\sign(\Delta\,\phi_1\phi_2)\,.\label{eq:mcn0}
\end{equation}
Therefore, starting from the trivial phase with $\Delta<0$ and creating a band inversion at \ptt{X} ($\Delta>0$) leads to a \mcn{}  of $\cn =2\, \sign (\phi_1\phi_2)$, where the factor $2$ comes from the fact that there are two \ptt{X}~points in the $k_x=0$ plane. 

The two other \mcn{}s can be calculated analogously, see \Ref{sm} for details. We obtain $\cp=1$ and $\cd=\nu \sign(\phi_1\phi_2)$, where $\nu=-1$ for a band inversion between \dxsys{} and a linear superposition of $\Gamma_8^{(1)}$ and $\Gamma_7$, and $\nu=1$ for a band inversion between \dzsrs{} and $\Gamma_8^{(2)}$.
Hence, if $\sign (\phi_1\phi_2)=1$ ($-1$), we recover the set of \mcn{}s which imply the pseudospin texture in \subfig{fig:spin-textures}{c} [\subfig{fig:spin-textures}{d}]. 
In general, we obtain 
\begin{align}
w_\pt{\bar X}=\sign(\phi_1\phi_2)\,. \label{eq:windphi12}
\end{align}

\mysection{Model calculations for \smb}\label{s:calc}
In the following, we will illustrate our theoretical findings by calculations with an effective lattice model for \smb. 
In the interest of simplicity, we will restrict ourselves to the 
$\Gamma_8$ quartet for $f$ electrons and study a model similar to that used in \Ref{takimoto_smb6_2011} and \Ref{baruselli_scanning_2014}. Analogous calculations can be performed for the full or the $\Gamma_7$ model~\mycite{sm}.
The Bloch Hamiltonian is an $8\times8$ matrix
\begin{subequations}\label{eq:gamma8model}%
\begin{align}%
	H_\mathrm{(\Gamma_8)}&= 
	\begin{pmatrix}
		h_d & {\Phi_8}^\dagger\\
		\Phi_8 & h_8
	\end{pmatrix}\,,
\end{align}
where the hopping of $d$ and $f$ electrons and the hybridization are given by
\begin{widetext}
	\allowdisplaybreaks
	\begin{align}
			h_d(\bk)&=\sigma_0
			\begin{pmatrix}
				-\frac{3}{2} (c_1+c_2) \left(t_d^{(1)}+2 t_d^{(2)} c_3\right) & \frac{\sqrt{3}}{2}  (c_1-c_2) \left(t_d^{(1)}-2 t_d^{(2)} c_3\right) \\
				\frac{\sqrt{3}}{2}  (c_1-c_2) \left(t_d^{(1)}-2 t_d^{(2)} c_3\right) & -4 t_d^{(2)} c_1 c_2-2 t_d^{(1)} c_3-\frac{1}{2} (c_1+c_2) \left(t_d^{(1)}+2 t_d^{(2)} c_3\right)
			\end{pmatrix}\,,\\
			h_{8}(\bk)&=\sigma_0 
			\begin{pmatrix}
				\epsilon_8-\frac{3}{2} (c_1+c_2) \left(t_8^{(1)}+2 t_8^{(2)} c_3\right) & \frac{ \sqrt{3}}{2} (c_1-c_2) \left(t_8^{(1)}-2 t_8^{(2)} c_3\right) \\
				\frac{ \sqrt{3}}{2} (c_1-c_2) \left(t_8^{(1)}-2 t_8^{(2)} c_3\right) & \epsilon_8-4 t_8^{(2)} c_1 c_2-2 t_8^{(1)} c_3-\frac{1}{2} (c_1+c_2) \left(t_8^{(1)}+2 t_8^{(2)} c_3\right)
			\end{pmatrix}\,,\\
			\Phi_8(\bk)&\!
			\begin{multlined}[t][.87\textwidth]
				=-\i\left(
				\begin{matrix}
					3/2\,V_8^{(1)}(s_1\sigma_1 + s_2\sigma_2)+3V_8^{(2)}\left[(c_1 + c_2)s_3\sigma_3 + c_3(s_1\sigma_1 + s_2\sigma_2)\right] \quad\dots\\
					-\sqrt{3}/2\,V_8^{(1)}(s_1\sigma_1 - s_2\sigma_2)+\sqrt{3}V_8^{(2)}\left[(c_1 - c_2)s_3\sigma_3 + c_3(s_1\sigma_1 - s_2\sigma_2)\right] \quad\dots
				\end{matrix}\right.\\
				\left.
				\begin{matrix}
					-\sqrt{3}/2\,V_8^{(1)}(s_1\sigma_1 - s_2\sigma_2)+\sqrt{3}V_8^{(2)}\left[(c_1 - c_2)s_3\sigma_3 + c_3(s_1\sigma_1 - s_2\sigma_2)\right]\\
					V_8^{(1)}\left[2s_3\sigma_3 + 1/2(s_1\sigma_1 + s_2\sigma_2)\right]+V_8^{(2)}\left[(c_1 + c_2)s_3\sigma_3 + 4(c_2s_1\sigma_1 + c_1s_2\sigma_2) + c_3(s_1\sigma_1 + s_2\sigma_2)\right]
				\end{matrix}\right)\,,
			\end{multlined}
			\label{eq:phi8}
		\end{align}
\end{widetext}%
\end{subequations}%
with the Pauli matrices $\sigma_\alpha$ acting in spin space and the spinor $\psi=(d_{x^2-y^2,\updown},d_{3z^2-r^2,\updown},f_{\Gamma_8^{(1)},\pm},f_{\Gamma_8^{(2)},\pm})\transp$.
Here, \mbox{$c_\alpha\equiv \cos k_\alpha$} and $s_\alpha\equiv \sin k_\alpha$, and we use $t^{(1,2)}$ ($V^{(1,2)}$) to denote first and second neighbor hopping (hybridization) parameters, respectively. 
The hopping and hybridization parameters should be considered as renormalized due to a strong local Coulomb interaction for the $f$ electrons~\mycite{Read:1983b,Coleman:1984,Rice:1985b}. As long as the electronic states near the Fermi energy are well described by quasiparticles, the adopted single-particle approach to compute the topological invariants is justified, even in the presence of strong electron correlations~\mycite{wang_simplified_2012,hohenadler_correlation_2013,werner_interaction_2013}.
A typical band structure in the \ptt{X}-inverted phase (without hybridization) is shown in \subfig{fig:st_gamma8}{a}.

\begin{figure}[tb]
	\renewcommand{\thisfig}{fig:st_gamma8}
	\centering
	
	\figlabel[0,.4]{\includegraphics[width=.22\textwidth]{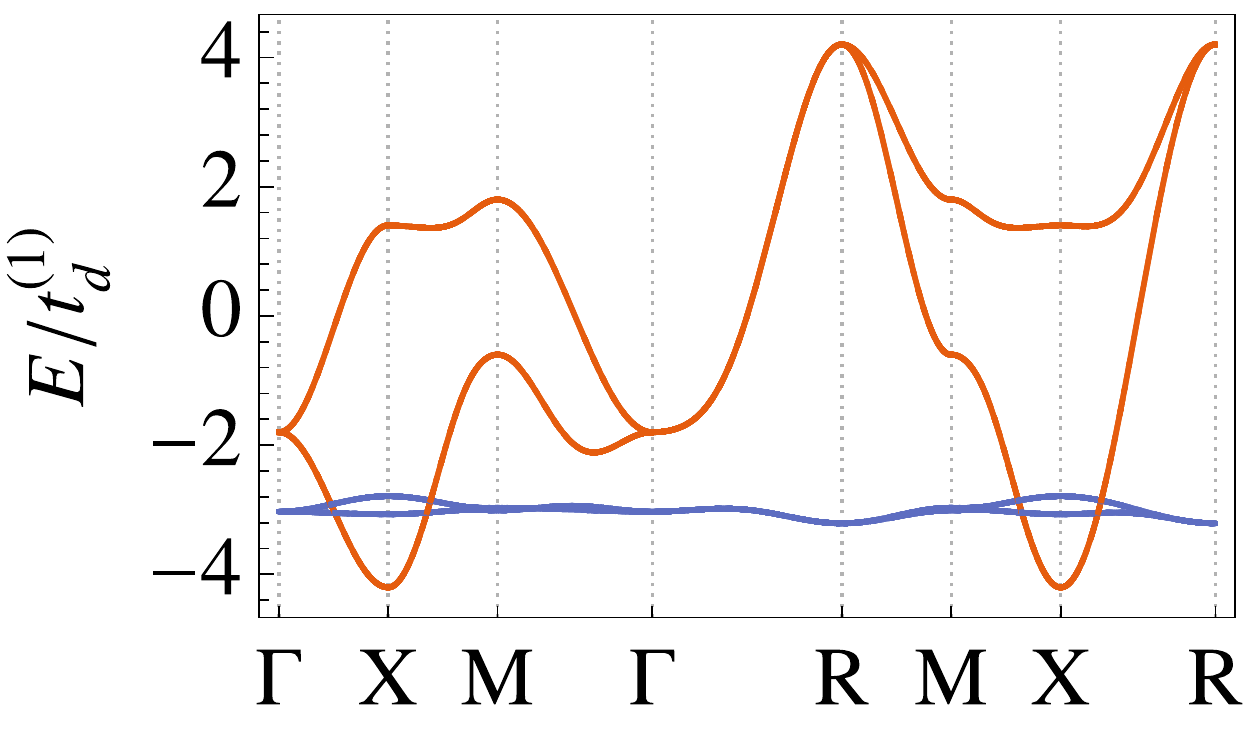}}{a}\hfill
	\figlabel[0,-.15]{\includegraphics{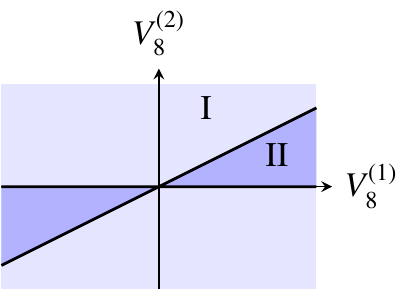}}{b}\\[5mm]
	
	\figlabel{\includegraphics[width=.22\textwidth]{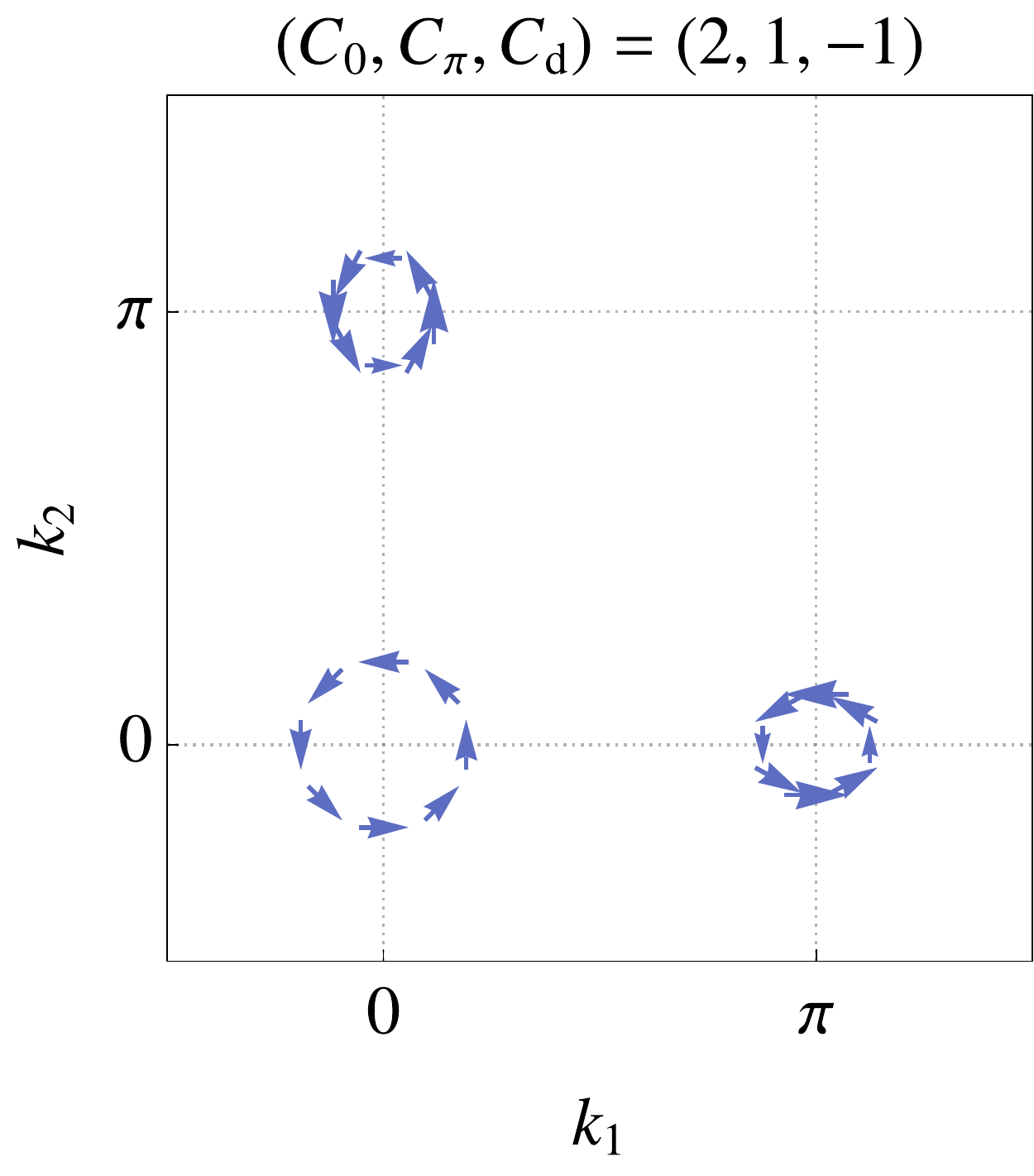}}{c}\hfill
	\figlabel{\includegraphics[width=.22\textwidth]{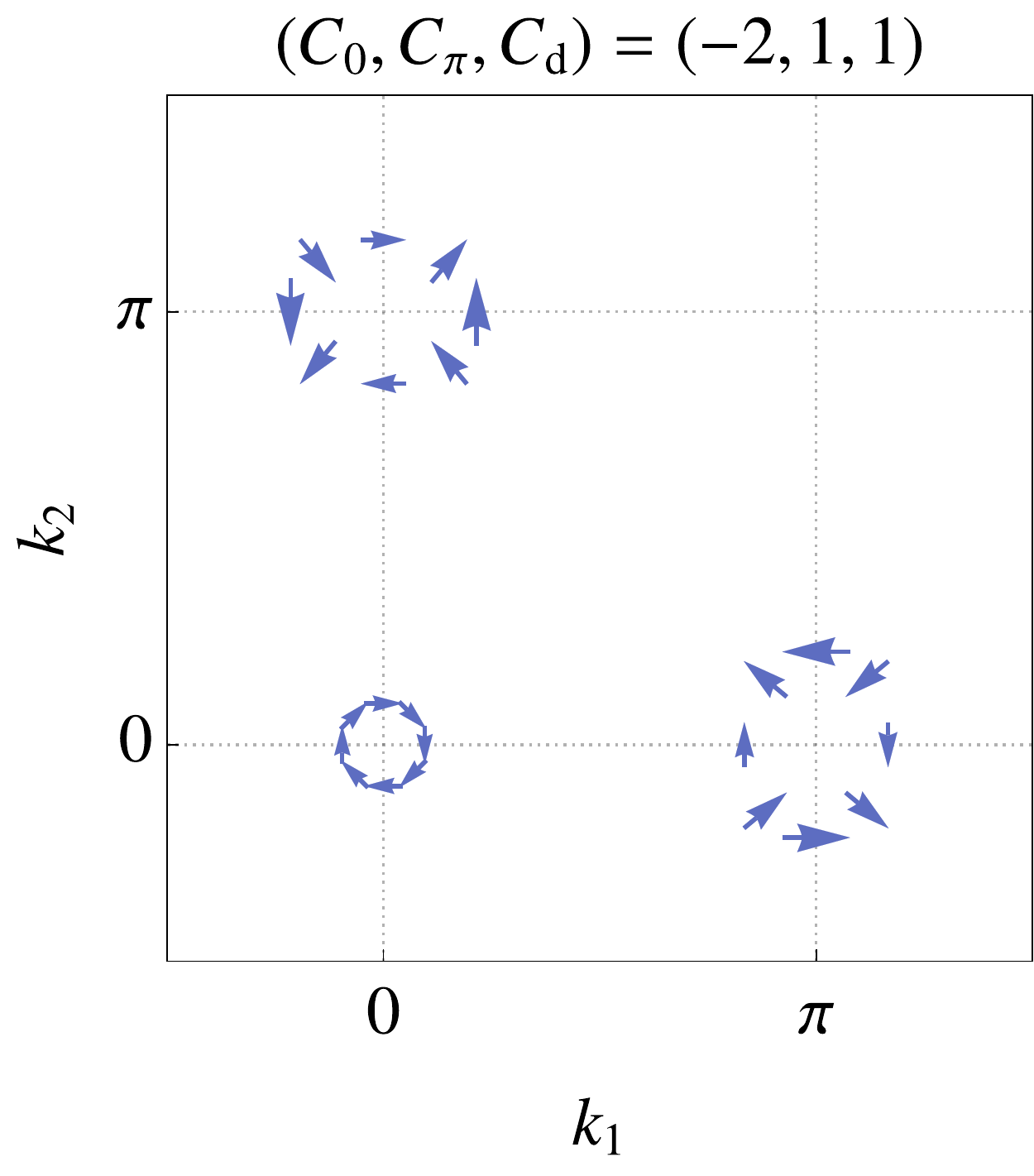}}{d}
	
	\caption{Band structure without hybridization (a) and phase diagram (b) for the $\Gamma_8$ model defined in \eq{eq:gamma8model} with $t_d^{(1)}=1$, $t_d^{(2)}=-0.2$, $t_8^{(1)}=-0.03$, $t_8^{(2)}=0.02$, and $\epsilon_8=-3$. The two spin textures (c-d) in phases I and II, respectively, are realized for the hybridization parameters  $(V_8^{(1)},V_8^{(2)})=(-0.1,0.1)$ and $(V_8^{(1)},V_8^{(2)})=(0.3,0.07)$, respectively.}
	\label{\thisfig}
\end{figure}

For $k_x=\pm k_y$, the offdiagonal elements of both $h_d$ and $h_8$ vanish and the $d$ and $f$ electrons are split into \dxsys{} and \dzsrs{} orbitals, and $\Gamma_8^{(1)}$ and $\Gamma_8^{(2)}$, respectively. 
Therefore, at the point $\pt{X}=(0,0,\pi)$ we obtain
\begin{equation}
	h_d(\pt{X})=\sigma_0\,\mathrm{diag}\left[-3\left(t_d^{(1)}-2t_d^{(2)}\right), t_d^{(1)}-2t_d^{(2)}\right]\,,
\end{equation}
and similarly for the $\Gamma_8$ orbitals.
\emph{Ab-initio} calculations~\mycite{deng_plutonium_2013} suggest that $t_d^{(1)},t_8^{(2)}>0$ and $t_d^{(2)},t_8^{(1)}<0$, such that the band inversion occurs between the \dxsys{} and the $\Gamma_8^{(1)}$ orbitals.
The hybridization matrix for these two orbitals can be expanded to first order at the \ptt{X}~point:
\begin{equation}\label{eq:hybX8}
	-\!\i\,{\Phi}_{\bs q}=\frac32 (\sigma_xq_x+\sigma_yq_y)\left(V_8^{(1)}-2V_8^{(2)}\right)\!-\!6V_8^{(2)}\sigma_zq_z\,.
\end{equation}
Therefore, according to \eq{eq:windphi12}, we obtain 
\begin{align}
\label{eq:C0}
	w_\pt{\bar X}=-\sign \left[ V_8^{(2)} \left(V_8^{(1)}-2V_8^{(2)}\right) \right]\,,
\end{align}
leading to the phase diagram shown in \subfig{fig:st_gamma8}{b}. 
As discussed \hyperref[s:mcn]{above}, $\nu=-1$ for \dxsys{} and $\Gamma_8^{(1)}$ orbitals, such that we expect $(\cn,\cp,\cd)=(2,1,-1)$ in phase I,  leading to a pseudospin texture with $w_{\pt{\bar X}}=1$, while we expect $(\cn,\cp,\cd)=(-2,1,1)$ and $w_{\pt{\bar X}}=-1$ in phase II. 
At the phase transitions $V_8^{(2)}=0$, the hybridization vanishes along the \ptt{\Gamma X} line, for $V_8^{(2)}=\frac12 V_8^{(1)}$ it vanishes at both the \ptt{XM} and \ptt{XR} lines.
This causes the hybridization gap to close and the \mcn{}s $(\cn,\cp,\cd)$ to change by $(\pm4,0,\mp2)$. 
We numerically confirmed the phase diagram in \subfig{fig:st_gamma8}{b} by directly calculating the \mcn{}s using a method for a discretized \bz~\mycite{suzuki_chern_2005}.
Figures~\subfigg{fig:st_gamma8}{c} and \ref{fig:st_gamma8:d} show the \emph{physical-spin} texture in phases I and II, respectively. 
They were calculated for a slab of 500 unit cells and fit the expected texture for the \emph{pseudospin}.

\mysection{Relation between physical spin and pseudospin}\label{s:spin}
The observed equivalence between physical-spin and pseudospin texture in \fig{fig:st_gamma8} requires more attention: Because the $f$ electrons experience strong spin-orbit coupling, the orbital pseudospin defined \hyperref[s:pseudospin]{above} is not equivalent to the physical spin of the electrons and the mirror and spin operators do not commute. The relation between physical and orbital pseudospin for the $J=5/2$ multiplet is given in \Ref{sm}.

According to the definition of the pseudospin \hyperref[s:pseudospin]{above}, a surface pseudospin in positive $\bs n$ direction corresponds to an eigenvalue $-\i$ of $M_{\bs n}$.
In order to find a relation between the physical-spin and pseudospin texture of the surface states, we therefore consider the effect of the projector $P^{\mathrm{ps}}_{\bs n}\equiv \frac12(\E+\i M_{\bs n})$ on the physical-spin operator $S_{\bs n}$, where $P^{\mathrm{ps}}_{\bs n}$ projects onto the subspace $M_{\bs n}=-\i$ and $\bs n$ is the normal vector of the mirror plane. One can show that, for the \Eg{} and $J=5/2$ multiplets,
\begin{align}
	P^{\mathrm{ps}}_{\bs n} S_{\bs n'} P^{\mathrm{ps}}_{\bs n}\equiv 0 \quad\text{for}\quad {\bs n}\perp{\bs n'}\,,\label{eq:spin_orth}
\end{align}
which states that on a \mil{}, the physical spin is always parallel (or antiparallel) to the surface-state pseudospin. Whether the two are parallel or antiparallel is determined by the eigenvalues of the projected spin operator,
\begin{align}\label{eq:spin_rest}
	S^{\mathrm{ps}}_{\bs n}\equiv P^{\mathrm{ps}}_{\bs n} S_{\bs n} P^{\mathrm{ps}}_{\bs n} \,.
\end{align}
For the $d$ orbitals we have $\bs S=\bs \sigma$ leading to eigenvalues $+1$ of $S^{\mathrm{ps}}_{\bs n}$, while for the $\Gamma_7$, the $\Gamma_8$, and the full model, we obtain the (approximate) spectra $\{-0.24\}$, $\{0.52, 0.14\}$, and $\{0.71, 0.14, -0.43\}$, respectively, see \Ref{sm}.
As all eigenvalues are positive for the $\Gamma_8$ model, the physical spin is indeed always \emph{parallel} to the surface-state pseudospin and all findings concerning the pseudospin are directly transferable to the physical spin.
This is not the case if we also consider the $\Gamma_7$ orbital, because the projected spin operator of $f$ electrons also has negative eigenvalues.
In these cases, the relation between pseudospin and physical spin of the surface states depends on the orbital character of the state.
In all cases we have studied, the \emph{winding number} of the physical spin sufficiently close to the Dirac node is nevertheless identical to the winding number of the pseudospin.
However, the direction may be reversed around some of the Dirac points.
Indeed, we find that this may occur for the $\Gamma_7$ model, signaling a dominant (in terms of spin) $\Gamma_7$ character of the surface states~\mycite{sm}. 

Finally, we mention that for other models with band crossings along some \hsl{}s, there is the possibility of phases with higher \mcn{}s and a larger number of protected surface states. We discuss an example in \Ref{sm}.

\mysection{Conclusion}
We have derived a close relationship between the hybridization matrix at the \ptt{X} high-symmetry points, the mirror Chern numbers, and the spin texture of the topologically protected surface states in topological Kondo insulators. 
Although we have motivated our study with \smb{}, the line of argumentation also applies to other topological insulators.
Explicit calculations for different models for \smb{} showed that the spin texture of the surface states does not only depend on the orbitals that are included in the effective model, but also depend on the magnitude of different hybridization parameters. 
This fact needs to be kept in mind when interpreting \emph{ab-initio} or effective-model-based calculations for this type of materials. 
Finally, our results can be used to infer mirror Chern numbers from spin-resolved \arpes{} measurements and predict further observables.

\mysection{Acknowledgments}
We would like to thank T.~Neupert, M.~Shi, and N.~Xu for inspiring discussions. 
This work is financially supported by a grant of the Swiss National Science Foundation. 

\mysection{Note added} During the submission process, a related study with compatible results has appeared~\mycite{baruselli_distinct_2015}.

% \vspace{2cm}
%-----------------------------------------------------%
%-----------------------------------------------------%
\mybibliography{Literatur_SmB6}
%merlin.mbs apsrev4-1.bst 2010-07-25 4.21a (PWD, AO, DPC) hacked
%Control: key (0)
%Control: author (8) initials jnrlst
%Control: editor formatted (1) identically to author
%Control: production of article title (-1) disabled
%Control: page (0) single
%Control: year (1) truncated
%Control: production of eprint (0) enabled
%

%-----------------------------------------------------%
%-----------------------------------------------------%
%-----------------------------------------------------%
\newcommand{\startsupplemental}{%
	\pagebreak
	\stepcounter{mypart}
	\setcounter{page}{1}
	\makeatletter
	\renewcommand{\theequation}{S\arabic{equation}}
	\renewcommand{\thepage}{S\arabic{page}}
	\renewcommand{\thefigure}{S\arabic{figure}}
	\renewcommand{\thetable}{S\Roman{table}}
	\renewcommand{\bibnumfmt}[1]{[S##1]}
	\renewcommand{\citenumfont}[1]{S##1}
	\onecolumngrid	
	\begin{center}
		\textbf{\large Supplemental Material for ``\mytitle''}
		\vspace{6mm}
	\end{center}
	\twocolumngrid
	\thispagestyle{empty}
	\renewcommand{\onlytwocol}[1]{##1}
	\renewcommand{\citeprefix}{SM_}
}
%-----------------------------------------------------%
\startsupplemental

\section{Mirror operators and sign choice of the mirror Chern numbers}

In general, the mirror operator for a plane with normal vector $\bs n$ can be written as
\begin{align}
	M_{\bs n}=IR_{\bs n}=I R^{\text{orb}}_{\bs n}R^{ \text{ps}}_{\bs n}\,,\label{eq:mirror_op-gen}
\end{align}
where $I$ is the inversion operator and $R_{\bs n}$ denotes a rotation by $\pi$ around $\bs n$. 
The spin part of the rotation is given by $R^{ \text{ps}}_{\bs n}=-\i(\bs n\cdot\bs \sigma)$, where $\bs \sigma$ are the Pauli matrices acting in spin space. $R^{\text{orb}}_{\bs n}$ denotes the orbital part of the rotation and depends on the symmetry of the considered orbitals. For example, for the orbitals $\{d_{x^2-y^2},d_{3z^2-r^2}, f_{\Gamma_7},f_{\Gamma_8^{(1)}},f_{\Gamma_8^{(2)}} \}$, the orbital part is $R^{\text{orb}}_{\bs n}=\E$ if $\bs n=\bs e_\alpha$ ($\alpha=x,y,z$); but for the mirror planes $k_\alpha=\pm k_\beta$ this operator contains additional nontrivial signs. Specifically, for $\bs n=\frac{1}{\sqrt{2}}(\bs e_y-\bs e_x)$ we obtain $R^{\text{orb}}_{\bs n}\psi_i=\nu_i\psi_i$ with $\nu=1$ in the subspaces $\{d_{3z^2-r^2}, f_{\Gamma_8^{(2)}} \}$ and $\nu = -1$ in the subspace $\{d_{x^2-y^2}, f_{\Gamma_8^{(1)}}, f_{\Gamma_7}\}$.

We note that the signs  of the \mcn{} are not uniquely determined in general:
One can choose the sign of the mirror operator $M$ and the orientation of the mirror invariant plane for the calculation of the \mcn{}, $\bs n_\mathrm{mp}$.
We fix the signs of the \mcn{}s by the convention
\begin{align}
	R^{ \text{ps}}_{\bs n_\mathrm{mp}}=-\i(\bs n_\mathrm{mp}\cdot\bs \sigma)\,.
\end{align}
The choice of orientation $\bs n_{mp}$ and the sign convention during the calculation of the Chern numbers still affects the positive directions in the \sbz{}. We use the convention where the Berry connection, Berry curvature, and Chern number are defined as $\mathcal A_\mu = \i \langle \psi(\bs k)|\partial_\mu|\psi(\bs k) \rangle$, $\mathcal F_{\mu\nu} = \partial_\mu \mathcal A_\nu - \partial_\nu \mathcal A_\mu$, and $C=\frac{1}{2\pi}\int \d k_1\d k_2 \mathcal F_{12}$, respectively. Note that this convention leads to an additional factor of $-1$ compared to \Ref{suzuki_chern_2005} which we use for our numerical calculations of the \mcn{}s.

For the given conventions for the definition of the mirror operator and the Chern number, the positive directions in the \sbz{} can be defined as
$
	\bs n_\mathrm{pos}=\bs n_\mathrm{sf}\times\bs n_\mathrm{mp} 
$,
where $\bs n_\mathrm{sf}$ is the outward pointing normal vector of the surface. 
In this work, we use the conventions $\bs n_\mathrm{mp}=\bs e_\alpha$ for a mirror-invariant plane $k_\alpha=0$ or $k_\alpha=\pi$  and $\bs n_\mathrm{mp}=\frac{1}{\sqrt{2}}(\bs e_y-\bs e_x)$ for the plane $k_x=k_y$.
This leads to the positive directions on the top ($\bs n_\mathrm{sf}=+\bs e_z$) (001) surface as shown in \subfig{fig:spin-textures}{b}.

\section{Mirror Chern number in the X-inverted phase}
In the Main Text we have calculated the \mcn{} $\cn$ from an expansion of the hybridization matrix around the \ptt{X} points. In the following we will explicitly show the analogous calculations for the two other \mcn{}s.

For the mirror plane $k_z=\pi$, the mirror operator is $M_z=-\i\tau_z\otimes\sigma_z$ and the effective Hamiltonian in the $M_z=+\i$ subspace is
\begin{equation}
H^{(+\i)}_{\mathrm{eff},k_z=\pi}({\bs q})=\bar{\varepsilon}\E+\phi_1(q_y\mu_x+q_x\mu_y)-\Delta\,\mu_z\,,
\end{equation}
using the basis vectors $(0,1,0,0)$ and $(0,0,1,0)$.
It only depends on the parameter $\phi_1$ and the contribution to the total Berry flux amounts to
\begin{equation}
C_{k_z=\pi}^\mathrm{Dirac}=\frac12\sign(\Delta\,\phi_1^2)=\frac12\sign(\Delta)\,.
\end{equation}
Using the same argument as for $\cn$ and with the fact that there is only one \ptt{X}~point in the $k_x=\pi$ plane, the \mcn{}  is therefore always $\cp =1$ in the \ptt{X}-inverted phase. 

Finally, consider the mirror plane $k_x=k_y$ with $M_{xy}=-\i \nu\tau_z\,\frac{1}{\sqrt2}(\sigma_y-\sigma_x)$, where $\nu=\pm 1$ is the orbital rotation eigenvalue in the considered subspace.
There, we obtain the effective Hamiltonian 
\begin{equation}
H^{(+\i)}_{\mathrm{eff},k_x=k_y}({\bs q})=\bar{\varepsilon}\E+\nu\phi_1q_{xy}\mu_x+\phi_2q_z\mu_y-\Delta\,\mu_z \,,
\end{equation}
where we use $q_{xy}:=\sqrt{2}q_x$ with $q_x=q_y$ and we chose the basis vectors $((1+\i)/2,\nu/\sqrt{2},0,0)$ and $(0,0,(1+\i)/2,-\nu /\sqrt{2})$.
The choice $\bs n_\mathrm{mp}=\frac{1}{\sqrt{2}}(\bs e_y-\bs e_x)$ corresponds to $k_1\equiv q_z$ and $k_2\equiv q_{xy}$.
Then, analogous to \eq{eq:mcn0}, we obtain the Berry flux
\begin{equation}
C_{k_x=k_y}^\mathrm{Dirac}=\frac12\nu \sign(\Delta\,\phi_1\phi_2)\,,
\end{equation}
corresponding to a \mcn{} $\cd =\nu \sign (\phi_1\phi_2)$ in the \ptt{X}-inverted phase.

\section{Simple model with NNN hybridization}	\label{app:a}
\renewcommand{\tt}{t}
\newcommand{\ttt}{t'}
\newcommand{\tttt}{t''}
\newcommand{\td}{t_{d}}
\newcommand{\tdd}{t_{d}'}
\newcommand{\tddd}{t_{d}''}
\newcommand{\tf}{t_{f}}
\newcommand{\tff}{t_{f}'}
\newcommand{\tfff}{t_{f}''}
\setlength{\tabcolsep}{3mm}

In \Ref{legner_topological_2014}, we have defined a simplified two-orbital model to describe \smb. In order to be able to discuss different spin textures of the surface states, we add a \nnn{} hybridization term to it.
Then, using the spinor $\psi=\left(d_\uparrow,d_\downarrow,f_\uparrow,f_\downarrow\right)\transp$, the Bloch Hamiltonian is given by
\begin{subequations}\label{eq:simplemodel}
\begin{equation}
H^{(\mathrm s)}(\bs k)=
\begin{pmatrix}h_d^{(\mathrm s)}(\bs k)&\Phi^{(\mathrm s)}(\bs k)\\\Phi^{(\mathrm s)}(\bs k)&h_f^{(\mathrm s)}(\bs k)\end{pmatrix}\,,
\label{eq:hsimple}
\end{equation}
where hopping and hybridization are defined as
\begin{widetext}
\allowdisplaybreaks
\begin{align}
h_d^{(\mathrm s)}(\bs k)&=\left[-2\td\, (c_1+c_2+c_3)-4\tdd\, (c_{12}+c_{23}+c_{31})-8\tddd\, c_1c_2c_3\right]\E \,,\label{eq:hd}\\
h_f^{(\mathrm s)}(\bs k)&=\left[\ef-2\tf (c_1+c_2+c_3)-4\tff (c_{12}+c_{23}+c_{31})-8\tfff c_1c_2c_3\right]\E \,,\label{eq:hf}\\
\Phi^{(\mathrm s)}(\bs k)&=-2\left[ \sigma_x s_x \left(V_1+V_2(c_y+c_z)\right)+\sigma_y s_y \left(V_1+V_2(c_z+c_x\vphantom{c_y})\right)+\sigma_z s_z \left(V_1+V_2(c_x+c_y)\right) \right]\,,\label{eq:phisimple}
\end{align}\label{eq:h_d_f_Hyb}%
\end{widetext}%
\end{subequations}%
with the definitions $c_\alpha\equiv\cos(k_\alpha)$, $c_{\alpha\beta}\equiv c_\alpha c_\beta$, and \mbox{$s_\alpha\equiv \sin(k_\alpha)$}. 
The band structure of this model without hybridization is shown in \subfig{fig:pd_simple}{a}.
In this simple model, the rotation operator is $R^{\text{orb}}_{\bs n}=\E$ for all mirror planes.
If there is only a \nn{} hybridization ($V_2=0$), then $\Phi(\bs k)=0$ is only possible at one of the \hsp{}s, $\bs k\in \{0,\pi\}^3$.
Therefore, only two possible sets of \mcn{}s are possible, see \tab{tab:mcnNN}.

\renewcommand{\arraystretch}{1.3}
\begin{table}[t]
	\caption{The two possible configurations (up to an overall minus sign) for \mcn{}s in the simple model with only \nn{} hybridization.
Multiple band inversions are additive for the \mcn{}s.
The first version has already been discussed in \Ref{legner_topological_2014}.}
	\begin{tabular}{c|cc}
		band inversion & $(\cn, \cp, \cd)_1$  & $(\cn, \cp, \cd)_2$\\\hline
		\g & (1,0,1) & (1,0,1)\\
		\x & (-2,1,-1) & (2,1,1)\\
		\m & (1,-2,-1) & (1,2,1)\\
		\r & (0,1,1) & (0,1,1)
	\end{tabular}
	\label{tab:mcnNN}
\end{table}

\begin{table}[t]
	\caption{Values for $V_1$ and $V_2$ and conditions for $\bs k$ such that $\Phi(\bs k)=0$ according to \eq{eq:phisimple} for the simple model (\hsp{}s are not shown). $k^*$ is a free parameter and permutations of the $k_\alpha$ are always allowed. Also shown are the \hsl{}s to which the shown values for $\bs k$ correspond.}
	\begin{tabular}{c|cc}
		$V_1$ & $\bs k$ & \hsl\\\hline
		$0$ & $(0,\pi,k^*)$ & \x\m\\
		$-2V_2$ & $(0,0,k^*)$ & \g\x\\
		$+2V_2$ & $(\pi,\pi,k^*)$ & \m\r\\
		$-V_2(1+\cos k^*)$ & $(0,k^*,\pm k^*)$ & \g\m\\
		$+V_2(1-\cos k^*)$ & $(\pi, k^*, \pm k^*)$ & \x\r\\
		$-2V_2 \cos k^*$ & $\bs k\in \{\pm k^*\}^3$ & \g\r
	\end{tabular}
	\label{tab:hybsimple}
\end{table}

However, in general there are multiple other possibilities for $\Phi(\bs k)=0$, if we allow for nonzero \nnn{} hybridization, see \tab{tab:hybsimple}.
If there exists a band crossing of the bare bands on one of the \hsl{}s, phase transitions are possible by changing the model parameters $V_1$ and $V_2$.
While the $\zz$ topological indices are invariant as no band inversion at the \hsp{}s is changed, the \mcn{}s can change with this procedure. 
This leads to a richer phase diagram which includes phases with higher values for the \mcn{}s and a large number of protected gapless surface modes. 
An example with two additional phases for fixed hopping parameters is shown in \subfig{fig:pd_simple}{b}.

The phase III exists due to the fact, that the gaps at the \ptt{XM} and \ptt{XR} lines do not close simultaneously as in the $\Gamma_7$ and $\Gamma_8$ models. 
The line separating it from phase II is obtained from calculating the crossing point of $d$- and $f$-electron bands along the \ptt{XR} line and using the formula from \tab{tab:hybsimple}. 
Similarly, the lines separating phases I and III are obtained by calculating the two crossings of the $d$ and $f$ bands along the \ptt{\Gamma M} lines.  
The two other lines, $V_2=-\frac12 V_1$ and $V_1=0$ correspond to gap closings at the \ptt{\Gamma X} and \ptt{XM} lines, respectively, similar to the results from the $\Gamma_8$ model in the Main Text. 
The \mcn{}s for all different phases were calculated numerically using the same method as in the Main Text~\mycite{suzuki_chern_2005}.

\begin{figure}[h]
	\renewcommand{\thisfig}{fig:pd_simple}
	\centering
	\figlabel[0,.4]{\includegraphics[width=.23\textwidth]{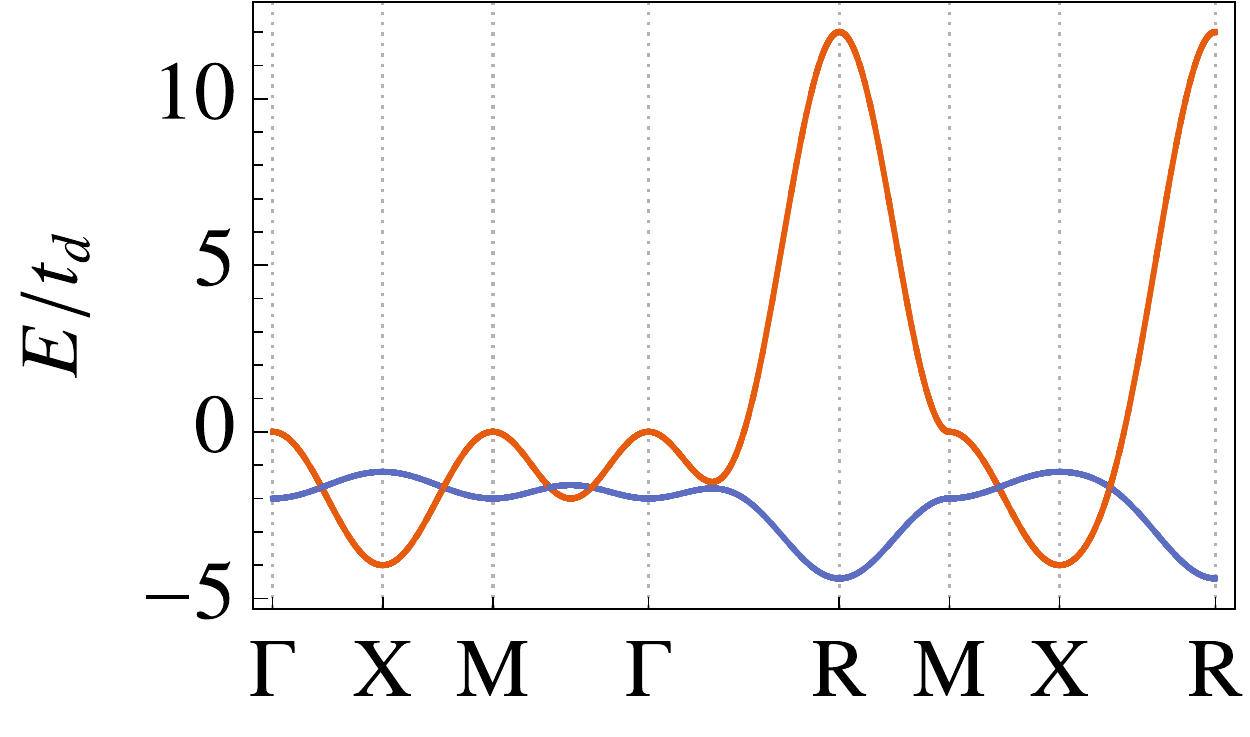}}{a}\onetwocol{\hspace{5mm}}{\hfill}
	\figlabel[0,-.15]{\includegraphics{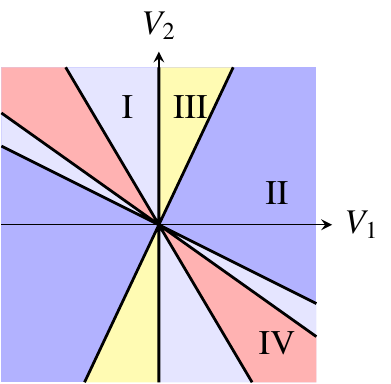}}{b}
	
	\caption{(a) Bandstructure without hybridization for the model defined in \eq{eq:simplemodel} with $t_d=1$, $t_d'=-0.5$, $t_d''=0$, $t_f=-0.2t_d$, and $\ef=-2$. The band crossing along the \ptt{\Gamma M} line combined with the fact that the hybridization can vanish along all \hsl{}s leads to a rich phase diagram (b). Phases I and II are the same as for the $\Gamma_8$ model in the Main Text. For the two additional phases III and IV, the \mcn{}s $(\cn,\cp,\cd)$ are $(-2,-3,1)$ and $(-2,1,3)$, respectively. Note that all phases are consistent with the general results of \Ref{ye_tci_2013}.}
	\label{\thisfig}
\end{figure}

\section{Realistic models for \texorpdfstring{\smb}{SmB6}}\label{app:b}
As discussed in the Main Text, for a more realistic description of \smb{}, usually the two $\rm E_g$ orbitals and one or several of the spin-orbit coupled $f$ orbitals $\Gamma_7$ and $\Gamma_8$ are used~\mycite{takimoto_smb6_2011, \citeprefix yu_model_2014, \citeprefix kim_termination_2014, \citeprefix baruselli_scanning_2014}.
The complete Hamiltonians can be constructed from the individual intra- and inter-orbital hopping amplitudes.
Here, we start from the model defined in \Ref{baruselli_scanning_2014} with a selection of nonzero parameters as a basis for our numerical calculations.
In the following definitions, $\tau_\alpha$ and $\sigma_\alpha$ will denote the Pauli matrices in orbital and spin space, respectively.
Also we will again use the definitions $c_\alpha\equiv\cos k_\alpha$ and $s_\alpha\equiv\sin k_\alpha$.

Then, with hopping and hybridization defined by
\begin{widetext}
	\allowdisplaybreaks
	\begin{subequations}\label{eq:hopping_hyb}%
		\begin{align}
			h_d(\bs k)&=\sigma_0
			\begin{pmatrix}
				-\frac{3}{2} (c_1+c_2) \left(t_d^{(1)}+2 t_d^{(2)} c_3\right) & \frac{\sqrt{3}}{2}  (c_1-c_2) \left(t_d^{(1)}-2 t_d^{(2)} c_3\right) \\
				\frac{\sqrt{3}}{2}  (c_1-c_2) \left(t_d^{(1)}-2 t_d^{(2)} c_3\right) & -4 t_d^{(2)} c_1 c_2-2 t_d^{(1)} c_3-\frac{1}{2} (c_1+c_2) \left(t_d^{(1)}+2 t_d^{(2)} c_3\right)
			\end{pmatrix}\,,\\
			h_{7}(\bs k)&=\sigma_0 \left[\epsilon_{7}-2t_7^{(1)}(c_1+c_2+c_3)-4t_7^{(2)}(c_1c_2+c_2c_3+c_3c_1)-8t_7^{(3)}c_1c_2c_3\right]\,,\\
			h_{8}(\bs k)&=\sigma_0 
			\begin{pmatrix}
				\epsilon_8-\frac{3}{2} (c_1+c_2) \left(t_8^{(1)}+2 t_8^{(2)} c_3\right) & \frac{ \sqrt{3}}{2} (c_1-c_2) \left(t_8^{(1)}-2 t_8^{(2)} c_3\right) \\
				\frac{ \sqrt{3}}{2} (c_1-c_2) \left(t_8^{(1)}-2 t_8^{(2)} c_3\right) & \epsilon_8-4 t_8^{(2)} c_1 c_2-2 t_8^{(1)} c_3-\frac{1}{2} (c_1+c_2) \left(t_8^{(1)}+2 t_8^{(2)} c_3\right)
			\end{pmatrix}\,,
		\end{align}
		\begin{align}
			h_{78}(\bs k)&\!
			\begin{multlined}[t][.87\textwidth]
			=\left(
					t_{78}^{(1)} \sigma_0\left(-2c_3+c_1+c_2\right)+2t_{78}^{(2)}\left((c_1+c_2) c_3+ -2c_1 c_2\right)\sigma_0+2\sqrt{3}\i t_{78}^{(2)} s_3(s_1\sigma_2 - s_2\sigma_1)
				\right. \quad \dots\\
				\hphantom{=\big(}\left. 
				\sqrt{3} t_{78}^{(1)} \sigma_0 (c_1-c_2)-t_{78}^{(2)}\left(2 \sqrt{3} \sigma_0 (c_1-c_2) c_3-4 \i \sigma_3 s_1 s_2+2 \i s_3 (\sigma_2 s_1+\sigma_1 s_2)\right)\right)\,,
			\end{multlined}\\
			\Phi_7(\bs k)&\!
			\begin{multlined}[t][.87\textwidth]\label{eq:phi7s}=-\i
				\left(V_7^{(1)}\left(2 s_3\sigma_3 -(s_1\sigma_1 +s_2\sigma_2 )\right)+  V_7^{(2)}\left(2 \sqrt{3} c_3 (s_1\sigma_1 +s_2\sigma_2 )-2 \sqrt{3} (c_1+c_2) s_3\sigma_3 \right)  \quad \dots \right.\\
				\left. V_7^{(1)}\left(-\sqrt{3} (s_1\sigma_1 -s_2\sigma_2 )\right)+  V_7^{(2)}\left(c_3 (s_1\sigma_1 -s_2\sigma_2 )+4 (c_2 s_1\sigma_1 -c_1 s_2\sigma_2 )-2 (c_1-c_2) s_3\sigma_3 \right) \right)\,,
			\end{multlined}\\
			\Phi_8(\bs k)&\!
			\begin{multlined}[t][.87\textwidth]
				=-\i\left(
				\begin{matrix}
					3/2\,V_8^{(1)}(s_1\sigma_1 + s_2\sigma_2)+3V_8^{(2)}\left[(c_1 + c_2)s_3\sigma_3 + c_3(s_1\sigma_1 + s_2\sigma_2)\right] \quad\dots\\
					-\sqrt{3}/2\,V_8^{(1)}(s_1\sigma_1 - s_2\sigma_2)+\sqrt{3}V_8^{(2)}\left[(c_1 - c_2)s_3\sigma_3 + c_3(s_1\sigma_1 - s_2\sigma_2)\right] \quad\dots
				\end{matrix}\right.\\
				\left.
				\begin{matrix}
					-\sqrt{3}/2\,V_8^{(1)}(s_1\sigma_1 - s_2\sigma_2)+\sqrt{3}V_8^{(2)}\left[(c_1 - c_2)s_3\sigma_3 + c_3(s_1\sigma_1 - s_2\sigma_2)\right]\\
					V_8^{(1)}\left[2s_3\sigma_3 + 1/2(s_1\sigma_1 + s_2\sigma_2)\right]+V_8^{(2)}\left[(c_1 + c_2)s_3\sigma_3 + 4(c_2s_1\sigma_1 + c_1s_2\sigma_2) + c_3(s_1\sigma_1 + s_2\sigma_2)\right]
				\end{matrix}\right)\,,
			\end{multlined}
		\end{align}%
	\end{subequations}%
\end{widetext}
and the spinor $\psi=(d_{x^2-y^2,\updown}, d_{3z^2-r^2,\updown}, f_{\Gamma_8^{(1)},\pm}, f_{\Gamma_8^{(2)},\pm}, f_{\Gamma_7\vphantom{\Gamma_8^(1)},\pm})\transp$, the Hamiltonian of the full model can be written as
	\begin{align}\label{eq:Hfull}
		H_\mathrm{full}&= 
		\begin{pmatrix}
			h_d & {\Phi_8}^\dagger & {\Phi_7}^\dagger\\
			\Phi_8 & h_8 & {h_{78}}^\dagger\\
			\Phi_7 & h_{78} & h_7
		\end{pmatrix}\,.
	\end{align}
Compared to \Ref{baruselli_scanning_2014}, our parameters are (for $\gamma=1,2,3$) $t_d^{(\gamma)}=\tilde t_d\eta_z^{d\gamma}$, $t_7^{(\gamma)}=\tilde t_f\eta_7^{f\gamma}$,  $t_8^{(\gamma)}=\tilde t_f \eta_z^{f\gamma}$, $t_{78}^{(\gamma)}=\tilde t_f\eta_{x7}^{f\gamma}$, $V_7^{(\gamma)}=\tilde v \eta_{z7}^{v\gamma}$, $V_8^{(\gamma)}=\tilde v \eta_{zz}^{v\gamma}$,  $\epsilon_7=\epsilon_{\Gamma_7\vphantom{\Gamma_8^(1)}}^f$, $\epsilon_8=\epsilon_{\Gamma_8}^f$, and we set $\epsilon^d=0$.

Reduced models including only the $\Gamma_7$ or only the $\Gamma_8$ orbitals can be obtained by removing one line and column of the Bloch matrix in \eq{eq:Hfull}.

\section{Calculations for the \texorpdfstring{$\Gamma_7$}{Gamma7} model}\label{s:gamma7}
Now we want to consider only the $\Gamma_7$ doublet for $f$ electrons and the \Eg{} quartet for $d$ electrons. Then, the Bloch Hamiltonian is given by
\begin{align}\label{eq:gamma7model}
	h_\mathrm{(\Gamma_7)}&= 
	\begin{pmatrix}
		h_d & {\Phi_7}^\dagger\\
		\Phi_7 & h_7
	\end{pmatrix}\,,
\end{align}
where the hopping and hybridization parts are defined in \eq{eq:hopping_hyb}.

\begin{figure}[bt]
	\renewcommand{\thisfig}{fig:st_gamma72}
	\centering
	\figlabel[0,.4]{\includegraphics[width=.22\textwidth]{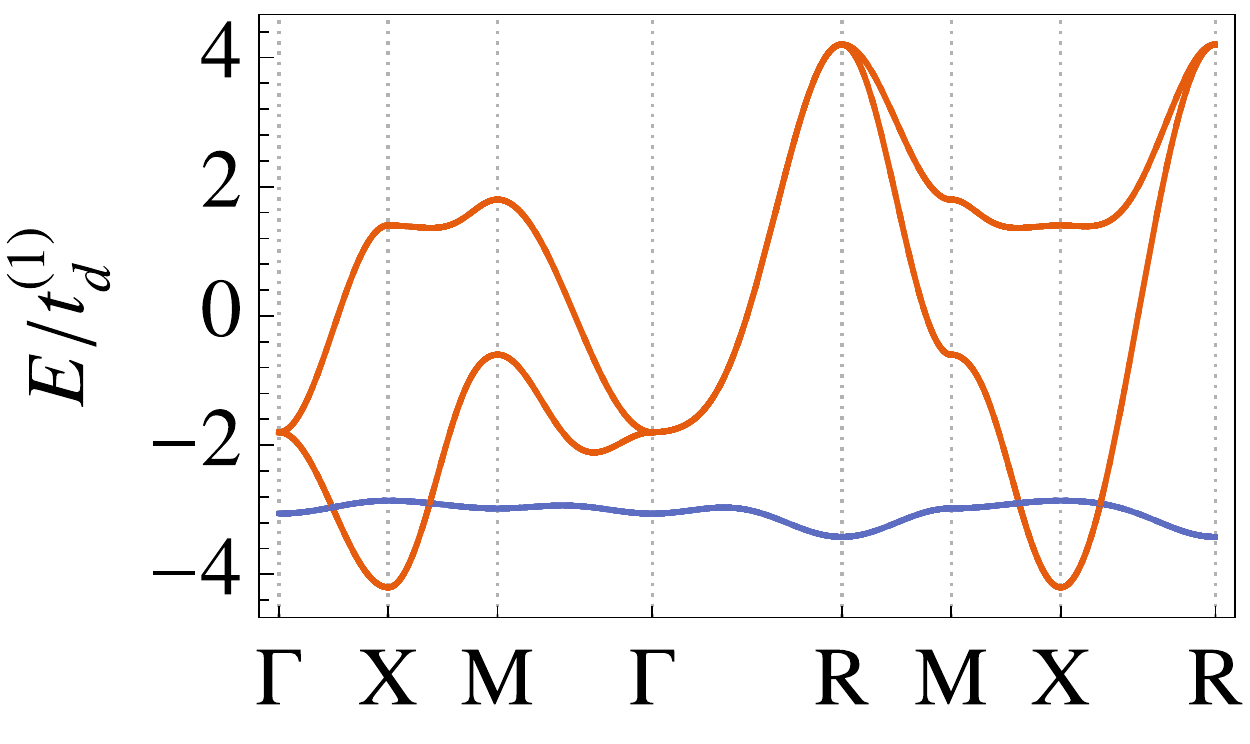}}{a}
	\onetwocol{\hspace{5mm}}{\hfill}
	\figlabel[0,-.15]{\includegraphics{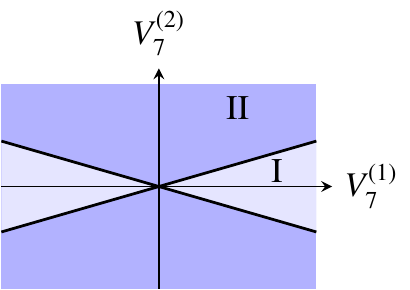}}{b}\\[5mm]
	\figlabel{\includegraphics[width=.22\textwidth]{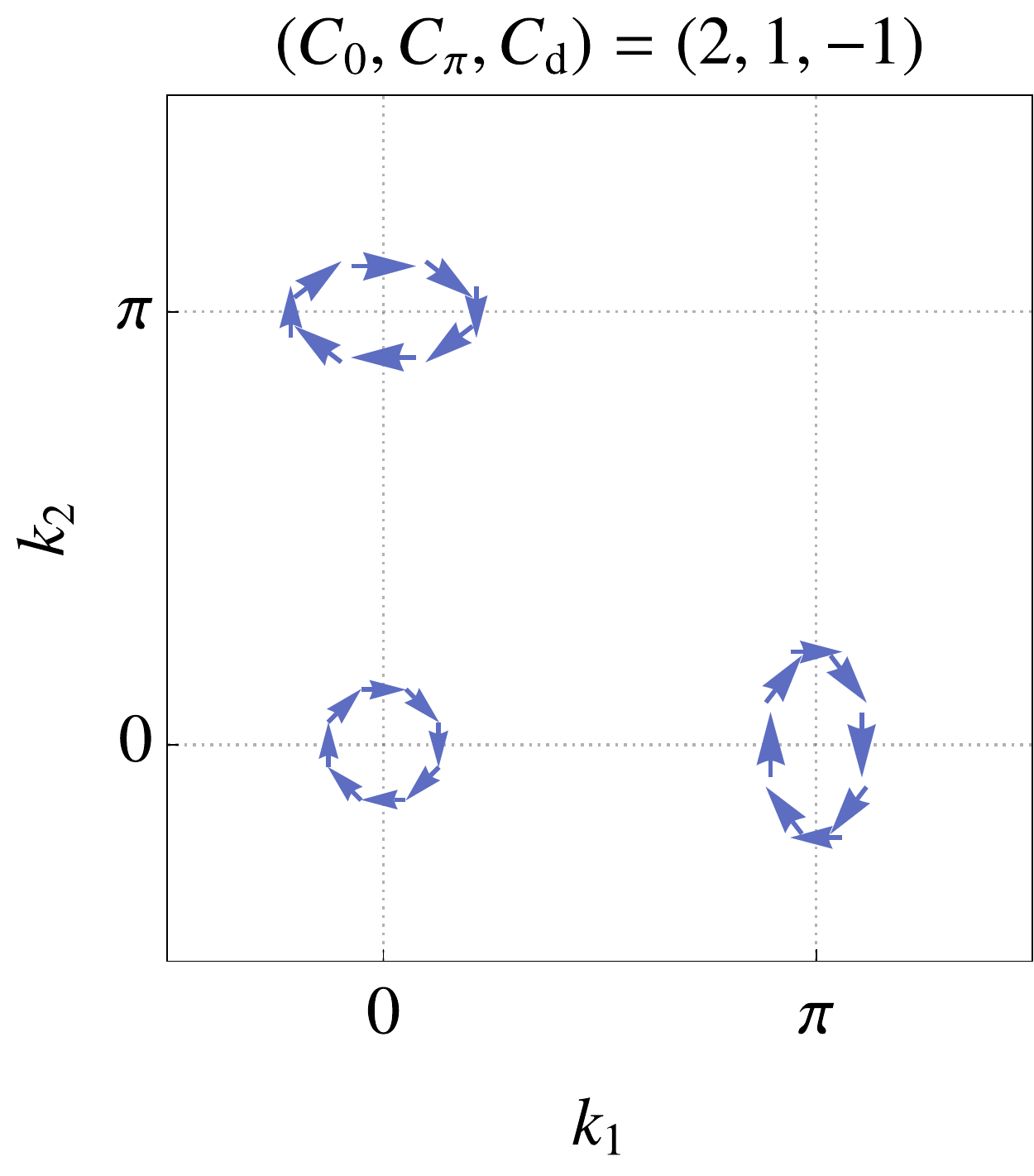}}{c}\onetwocol{\hspace{5mm}}{\hfill}
	\figlabel{\includegraphics[width=.22\textwidth]{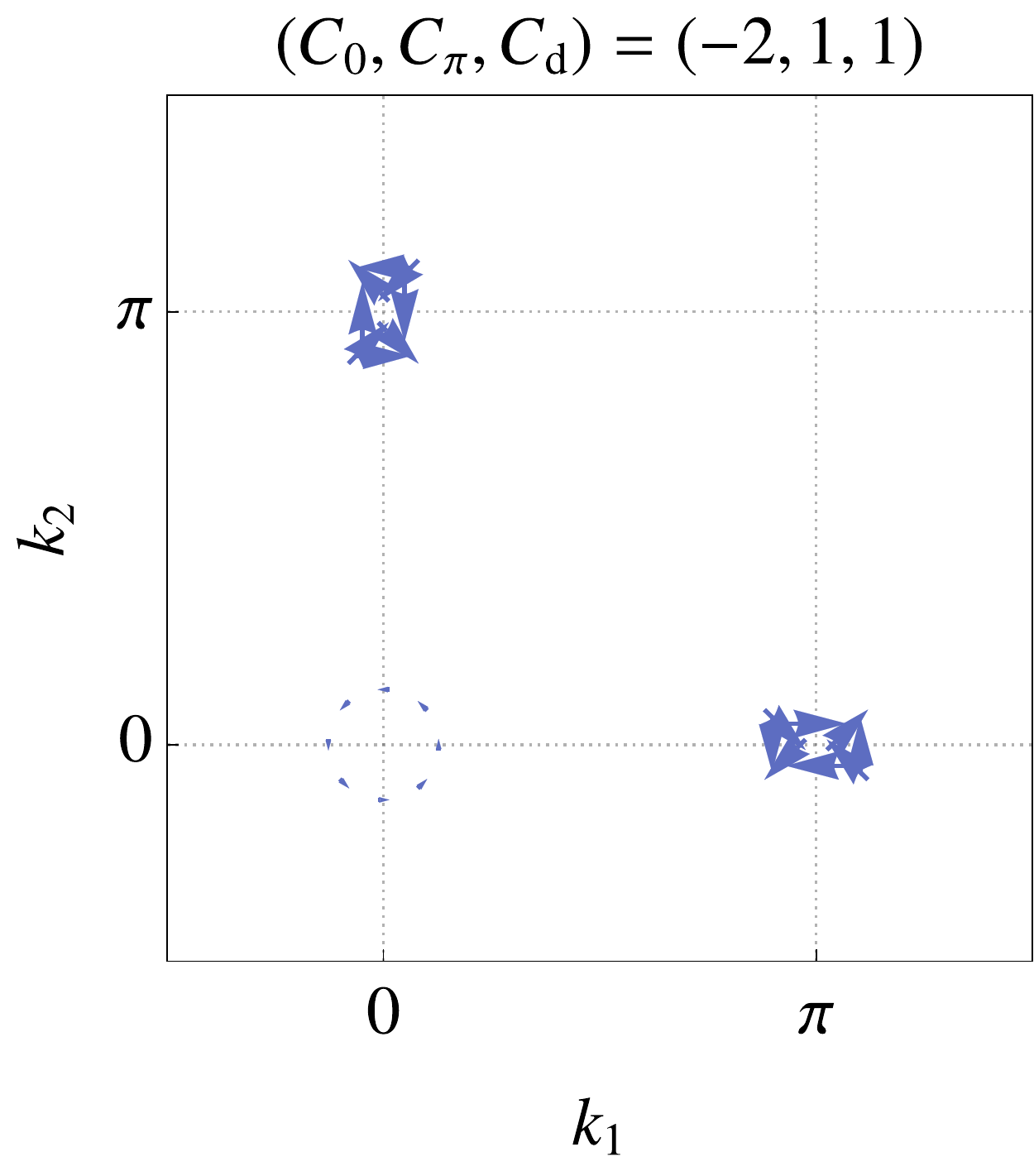}}{d}
	
	\caption{Bandstructure without hybridization (a) and phase diagram (b) for the $\Gamma_7$ model defined in \eq{eq:gamma7model} and \eqref{eq:hopping_hyb} with $t_d^{(1)}=1$, $t_d^{(2)}=-0.2$, $t_7^{(1)}=-0.03$, $t_7^{(2)}=0.02$, and $\epsilon_7=-3$. The two different spin textures (c-d) in phases I and II, respectively, are shown for hybridization parameters $(V_7^{(1)},V_7^{(2)})=(0.3,0)$ and $(V_7^{(1)},V_7^{(2)})=(0.1,0.1)$, respectively. Due to the negative eigenvalue of the restricted spin operator for $f$ electrons (\eq{eq:spin_rest}), the spin direction is reversed around all \hsp{}s when compared to \fig{fig:spin-textures}. Note that the magnitude of the spin expectation value around the \ptt{\bar\Gamma} is relatively small in (b) due to the mixing of $f$ and $d$ orbitals.}
	\label{\thisfig}
\end{figure}

As in the Main Text, we consider the situation where the \dxsys{} orbitals are lower in energy at the point $\pt{X}=(0,0,\pi)$ such that the inversion occurs between those and the $\Gamma_7$ orbitals. 
The hybridization matrix for these two orbitals can be expanded to first order at the \ptt{X}~point:
\begin{multline}\label{eq:hybX7}
	-\!\i\,{\Phi}_{\bs q}=(\sigma_xq_x+\sigma_yq_y)\left(V_7^{(1)}+2\sqrt3 V_7^{(2)}\right)\onetwocol{}{\\}+\sigma_zq_z\left(2V_7^{(1)}-4\sqrt3 V_7^{(2)}\right)\,.
\end{multline}
Therefore, we obtain 
\begin{align}
	\cn=2\sign\left[\big(V_7^{(1)}\big)^2-12\big(V_7^{(2)}\big)^2\right]\,,
\end{align}
leading to the phase diagram shown in \subfig{fig:st_gamma72}{b}. 
For dominant nearest-neighbor hybridization (phase I) we therefore expect a spin texture with $w_{\pt{\bar X}}=1$, while we expect $w_{\pt{\bar X}}=-1$ for dominant next-to-nearest neighbor hybridization (phase II), see also \fig{fig:spin-textures}. 
At the phase transition $V_7^{(1)}=2\sqrt{3}V_7^{(2)}$, the hybridization vanishes along the \ptt{\Gamma X} line, for $V_7^{(1)}=-2\sqrt{3}V_7^{(2)}$ at both the \ptt{XM} and \ptt{XR} lines, as for the $\Gamma_8$ model in the Main Text.
Our numerical calculations for the $\Gamma_7$ model confirm both the \mcn{}s and the expected spin texture, see \subfigs{fig:st_gamma72}{c}{d}.

\section{Calculations for the full model}
Assuming, as in the Main Text, that $t_d^{(1)},t_8^{(2)}>0$ and $t_d^{(2)},t_8^{(1)}<0$, the \mcn{}s and therefore the spin texture depend on the energy of $\Gamma_7$   and $\Gamma_8^{(1)}$ orbitals at the \ptt{X} point, $\varepsilon_7^\pt{X}$ and $\varepsilon_8^\pt{X}$, respectively.
If the energy difference $\Delta\varepsilon^\pt{X} \equiv \varepsilon_8^\pt{X}-\varepsilon_7^\pt{X}$ is much larger than the hopping between $\Gamma_7$ and $\Gamma_8$ orbitals, $\Delta\varepsilon^\pt{X} \gg t_{78}\equiv \max(|t_{78}^{(1)}|,|t_{78}^{(2)}|)$, the band inversion essentially occurs between the \dxsys{} and the $\Gamma_8^{(1)}$ orbitals.
Then, hopping and hybridization of the $\Gamma_7$ orbitals is irrelevant for the topology and all results for the $\Gamma_8$ model in the Main Text are applicable.
Instead, for $\Delta\varepsilon^\pt{X} \ll -|t_{78}^{(1)}|$, the $\Gamma_8^{(1)}$ orbitals are not involved in the band inversion such that we can apply the results of the $\Gamma_7$ model.

However, at the \ptt{X} point there exists a non-vanishing mixing between those two orbitals, which we need to take into account for comparable energies of $\Gamma_7$ and $\Gamma_8^{(1)}$ orbitals. 
This leads to a continuous crossover between the two situations.
For $\Delta\varepsilon^\pt{X}=0$, we obtain an avoided crossing at the \ptt{X} point with the splitting being defined by $t_{78}^{(1)}$ and $t_{78}^{(2)}$. Then, the highest $f$ band is an equal superposition of $\Gamma_7$ and $\Gamma_8^{(1)}$ orbitals, 
\begin{equation}\label{eq:superpos78}
\psi^\pt{X}_\pm=\frac{1}{\sqrt{2}}\left(f_{\Gamma_7\vphantom{\Gamma_8^(1)}}^\pt{X} \pm f_{\Gamma_8^{(1)}}^\pt{X}\right)\,,
\end{equation}
where the sign $\pm$ depends on the parameters $t_{78}^{(1)}$ and $t_{78}^{(2)}$.

\begin{figure}[tb]
	\renewcommand{\thisfig}{fig:pd_full}
	\centering
	\figlabel[0,-.15]{\includegraphics{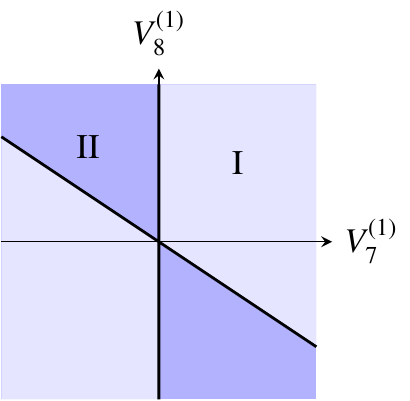}}{a}
	\onetwocol{\hspace{5mm}}{\hfill}
	\figlabel[0,-.15]{\includegraphics{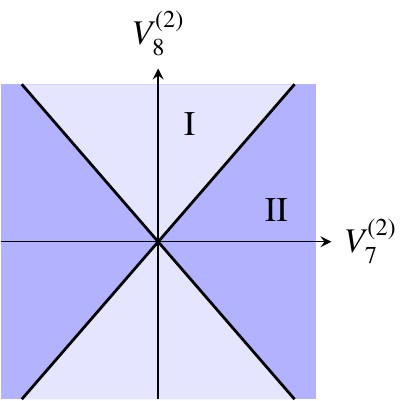}}{b}
	
	\caption{Phase diagrams for the full model in the case where $\psi_+^\pt{X}$ defined in \eq{eq:superpos78} is the highest $f$ orbital and $V_7^{(2)}=V_8^{(2)}=0$ (a) and $V_7^{(1)}=V_8^{(1)}=0$ (b). The phases I and II correspond to $w_\pt{\bar X}=1$ and $w_\pt{\bar X}=-1$, respectively.}
	\label{\thisfig}
\end{figure}

For this orbital, the hybridization matrix with the \dxsys{} at the \ptt{X} point to first order is given by a superposition of the matrices defined in Eqs.~\eqref{eq:hybX8} and \eqref{eq:hybX7}:
\begin{multline}
	-\!\i\sqrt2\,\Phi_{\bs q} = (\sigma_xq_x+\sigma_yq_y)\left(V_7^{(1)}+2\sqrt3 V_7^{(2)}\pm \tfrac32 V_8^{(1)} \mp 3V_8^{(2)}\right)\onetwocol{}{\\}
	+\sigma_zq_z\left(2V_7^{(1)}-4\sqrt3 V_7^{(2)} \mp 6V_8^{(2)}\right)\,.
\end{multline}
Then, the winding number of the spin texture at the \ptt{\bar X} points depends on all four hybridization parameters,
\begin{multline}
	w_{\pt{\bar X}} = \sign\left[\left(V_7^{(1)}+2\sqrt3 V_7^{(2)}\pm \tfrac32 V_8^{(1)} \mp 3V_8^{(2)}\right) \onetwocol{}{\right.\\\left. \cdot}
	\left(V_7^{(1)}-2\sqrt3 V_7^{(2)} \mp 3V_8^{(2)}\right)\right] \,.
\end{multline}
This is a complicated phase diagram where the two different phases are separated by two hyper planes. 
For illustration purposes, we show the phase diagrams for the positive sign in \eq{eq:superpos78}, and only nearest and only next-to-nearest neighbor hybridizations in \subfigs{fig:pd_full}{a}{b}, respectively.

\section{Spin operator for \texorpdfstring{$f$}{f} orbitals}\label{app:c}
While the spin operator for the $d$ orbitals is a block-diagonal matrix $ {\bs S}^d=\E \otimes \bs\sigma$, this is not the case for the $f$ orbitals~\mycite{baruselli_scanning_2014}.
The spin operator for the $f$ orbitals can be obtained by calculating its matrix elements for $\Gamma_7$ and $\Gamma_8$ states.
Writing the spinor of $f$ electrons as $\psi=\Big(f_{\Gamma_8^{(1)},\pm},f_{\Gamma_8^{(2)},\pm}, f_{\Gamma_7\vphantom{\Gamma_8^(1)}}\Big)\transp$, we obtain the spin operators
\begin{subequations}
\begin{align}
	 S^f_x\onetwocol{}{&}=
	\frac{1}{21}\,\sigma_x\begin{pmatrix}
		-5&-2\sqrt{3}&2\sqrt{5}\\
		-2\sqrt{3}&-9&2\sqrt{15}\\
		2\sqrt{5}&2\sqrt{15}&5
	\end{pmatrix}
	\,, \onetwocol{\quad}{\\}
	 S^f_y\onetwocol{}{&}=
	\frac{1}{21}\,\sigma_y\begin{pmatrix}
		-5&2\sqrt{3}&2\sqrt{5}\\
		2\sqrt{3}&-9&-2\sqrt{15}\\
		2\sqrt{5}&-2\sqrt{15}&5
	\end{pmatrix}
	\,, \onetwocol{\quad}{\\}
	 S^f_z\onetwocol{}{&}=\frac{1}{21}\,\sigma_z
	\begin{pmatrix}
		-11&0&-4\sqrt{5}\\
		0&-3&0\\
		-4\sqrt{5}&0&5
	\end{pmatrix}\,.
\end{align}
\end{subequations}
Again, the spin matrix for the reduced models is obtained by removing the matrix elements that include either the $\Gamma_7$ or $\Gamma_8$ orbitals.

%-----------------------------------------------------%
%-----------------------------------------------------%
\mybibliography{Literatur_SmB6}


\begin{thebibliography}{46}%
\makeatletter
\providecommand \@ifxundefined [1]{%
 \@ifx{#1\undefined}
}%
\providecommand \@ifnum [1]{%
 \ifnum #1\expandafter \@firstoftwo
 \else \expandafter \@secondoftwo
 \fi
}%
\providecommand \@ifx [1]{%
 \ifx #1\expandafter \@firstoftwo
 \else \expandafter \@secondoftwo
 \fi
}%
\providecommand \natexlab [1]{#1}%
\providecommand \enquote  [1]{``#1''}%
\providecommand \bibnamefont  [1]{#1}%
\providecommand \bibfnamefont [1]{#1}%
\providecommand \citenamefont [1]{#1}%
\providecommand \href@noop [0]{\@secondoftwo}%
\providecommand \href [0]{\begingroup \@sanitize@url \@href}%
\providecommand \@href[1]{\@@startlink{#1}\@@href}%
\providecommand \@@href[1]{\endgroup#1\@@endlink}%
\providecommand \@sanitize@url [0]{\catcode `\\12\catcode `\$12\catcode
  `\&12\catcode `\#12\catcode `\^12\catcode `\_12\catcode `\%12\relax}%
\providecommand \@@startlink[1]{}%
\providecommand \@@endlink[0]{}%
\providecommand \url  [0]{\begingroup\@sanitize@url \@url }%
\providecommand \@url [1]{\endgroup\@href {#1}{\urlprefix }}%
\providecommand \urlprefix  [0]{URL }%
\providecommand \Eprint [0]{\href }%
\providecommand \doibase [0]{http://dx.doi.org/}%
\providecommand \selectlanguage [0]{\@gobble}%
\providecommand \bibinfo  [0]{\@secondoftwo}%
\providecommand \bibfield  [0]{\@secondoftwo}%
\providecommand \translation [1]{[#1]}%
\providecommand \BibitemOpen [0]{}%
\providecommand \bibitemStop [0]{}%
\providecommand \bibitemNoStop [0]{.\EOS\space}%
\providecommand \EOS [0]{\spacefactor3000\relax}%
\providecommand \BibitemShut  [1]{\csname bibitem#1\endcsname}%
\let\auto@bib@innerbib\@empty
%</preamble>
\bibitem [{\citenamefont {Dzero}\ \emph {et~al.}(2010)\citenamefont {Dzero},
  \citenamefont {Sun}, \citenamefont {Galitski},\ and\ \citenamefont
  {Coleman}}]{dzero_topological_2010}%
  \BibitemOpen
  \bibfield  {author} {\bibinfo {author} {\bibfnamefont {M.}~\bibnamefont
  {Dzero}}, \bibinfo {author} {\bibfnamefont {K.}~\bibnamefont {Sun}}, \bibinfo
  {author} {\bibfnamefont {V.}~\bibnamefont {Galitski}}, \ and\ \bibinfo
  {author} {\bibfnamefont {P.}~\bibnamefont {Coleman}},\ }\href
  {http://link.aps.org/doi/10.1103/PhysRevLett.104.106408} {\bibfield
  {journal} {\bibinfo  {journal} {Phys.\ Rev.\ Lett.}\ }\textbf {\bibinfo
  {volume} {104}},\ \bibinfo {pages} {106408} (\bibinfo {year}
  {2010})}\BibitemShut {NoStop}%
\bibitem [{\citenamefont {Dzero}\ \emph {et~al.}(2012)\citenamefont {Dzero},
  \citenamefont {Sun}, \citenamefont {Coleman},\ and\ \citenamefont
  {Galitski}}]{dzero_theory_2012}%
  \BibitemOpen
  \bibfield  {author} {\bibinfo {author} {\bibfnamefont {M.}~\bibnamefont
  {Dzero}}, \bibinfo {author} {\bibfnamefont {K.}~\bibnamefont {Sun}}, \bibinfo
  {author} {\bibfnamefont {P.}~\bibnamefont {Coleman}}, \ and\ \bibinfo
  {author} {\bibfnamefont {V.}~\bibnamefont {Galitski}},\ }\href
  {http://link.aps.org/doi/10.1103/PhysRevB.85.045130} {\bibfield  {journal}
  {\bibinfo  {journal} {Phys.\ Rev.\ B}\ }\textbf {\bibinfo {volume} {85}},\
  \bibinfo {pages} {045130} (\bibinfo {year} {2012})}\BibitemShut {NoStop}%
\bibitem [{\citenamefont {Takimoto}(2011)}]{takimoto_smb6_2011}%
  \BibitemOpen
  \bibfield  {author} {\bibinfo {author} {\bibfnamefont {T.}~\bibnamefont
  {Takimoto}},\ }\href {\doibase 10.1143/JPSJ.80.123710} {\bibfield  {journal}
  {\bibinfo  {journal} {J.\ Phys.\ Soc.\ Jpn.}\ }\textbf {\bibinfo {volume}
  {80}},\ \bibinfo {pages} {123710} (\bibinfo {year} {2011})}\BibitemShut
  {NoStop}%
\bibitem [{\citenamefont {Tran}\ \emph {et~al.}(2012)\citenamefont {Tran},
  \citenamefont {Takimoto},\ and\ \citenamefont {Kim}}]{tran_phase_2012}%
  \BibitemOpen
  \bibfield  {author} {\bibinfo {author} {\bibfnamefont {M.-T.}\ \bibnamefont
  {Tran}}, \bibinfo {author} {\bibfnamefont {T.}~\bibnamefont {Takimoto}}, \
  and\ \bibinfo {author} {\bibfnamefont {K.-S.}\ \bibnamefont {Kim}},\ }\href
  {\doibase 10.1103/PhysRevB.85.125128} {\bibfield  {journal} {\bibinfo
  {journal} {Phys. Rev. B}\ }\textbf {\bibinfo {volume} {85}},\ \bibinfo
  {pages} {125128} (\bibinfo {year} {2012})}\BibitemShut {NoStop}%
\bibitem [{\citenamefont {Lu}\ \emph {et~al.}(2013)\citenamefont {Lu},
  \citenamefont {Zhao}, \citenamefont {Weng}, \citenamefont {Fang},\ and\
  \citenamefont {Dai}}]{lu_correlated_2013}%
  \BibitemOpen
  \bibfield  {author} {\bibinfo {author} {\bibfnamefont {F.}~\bibnamefont
  {Lu}}, \bibinfo {author} {\bibfnamefont {J.~Z.}\ \bibnamefont {Zhao}},
  \bibinfo {author} {\bibfnamefont {H.}~\bibnamefont {Weng}}, \bibinfo {author}
  {\bibfnamefont {Z.}~\bibnamefont {Fang}}, \ and\ \bibinfo {author}
  {\bibfnamefont {X.}~\bibnamefont {Dai}},\ }\href {\doibase
  10.1103/PhysRevLett.110.096401} {\bibfield  {journal} {\bibinfo  {journal}
  {Phys. Rev. Lett.}\ }\textbf {\bibinfo {volume} {110}},\ \bibinfo {pages}
  {096401} (\bibinfo {year} {2013})}\BibitemShut {NoStop}%
\bibitem [{\citenamefont {Alexandrov}\ \emph {et~al.}(2013)\citenamefont
  {Alexandrov}, \citenamefont {Dzero},\ and\ \citenamefont
  {Coleman}}]{alexandrov_cubic_2013}%
  \BibitemOpen
  \bibfield  {author} {\bibinfo {author} {\bibfnamefont {V.}~\bibnamefont
  {Alexandrov}}, \bibinfo {author} {\bibfnamefont {M.}~\bibnamefont {Dzero}}, \
  and\ \bibinfo {author} {\bibfnamefont {P.}~\bibnamefont {Coleman}},\ }\href
  {\doibase 10.1103/PhysRevLett.111.226403} {\bibfield  {journal} {\bibinfo
  {journal} {Phys.\ Rev.\ Lett.}\ }\textbf {\bibinfo {volume} {111}},\ \bibinfo
  {pages} {226403} (\bibinfo {year} {2013})}\BibitemShut {NoStop}%
\bibitem [{\citenamefont {Dzero}\ and\ \citenamefont
  {Galitski}(2013)}]{dzero_new_2013}%
  \BibitemOpen
  \bibfield  {author} {\bibinfo {author} {\bibfnamefont {M.}~\bibnamefont
  {Dzero}}\ and\ \bibinfo {author} {\bibfnamefont {V.}~\bibnamefont
  {Galitski}},\ }\href {\doibase 10.1134/S1063776113110083} {\bibfield
  {journal} {\bibinfo  {journal} {J.\ Exp.\ Theor.\ Phys.}\ }\textbf {\bibinfo
  {volume} {117}},\ \bibinfo {pages} {499} (\bibinfo {year}
  {2013})}\BibitemShut {NoStop}%
\bibitem [{\citenamefont {{Ye}}\ \emph {et~al.}(2013)\citenamefont {{Ye}},
  \citenamefont {{Allen}},\ and\ \citenamefont {{Sun}}}]{ye_tci_2013}%
  \BibitemOpen
  \bibfield  {author} {\bibinfo {author} {\bibfnamefont {M.}~\bibnamefont
  {{Ye}}}, \bibinfo {author} {\bibfnamefont {J.~W.}\ \bibnamefont {{Allen}}}, \
  and\ \bibinfo {author} {\bibfnamefont {K.}~\bibnamefont {{Sun}}},\
  }\href@noop {} {\bibfield  {journal} {\bibinfo  {journal} {ArXiv e-prints}\ }
  (\bibinfo {year} {2013})},\ \Eprint {http://arxiv.org/abs/1307.7191}
  {arXiv:1307.7191 [cond-mat.str-el]} \BibitemShut {NoStop}%
\bibitem [{\citenamefont {Deng}\ \emph {et~al.}(2013)\citenamefont {Deng},
  \citenamefont {Haule},\ and\ \citenamefont {Kotliar}}]{deng_plutonium_2013}%
  \BibitemOpen
  \bibfield  {author} {\bibinfo {author} {\bibfnamefont {X.}~\bibnamefont
  {Deng}}, \bibinfo {author} {\bibfnamefont {K.}~\bibnamefont {Haule}}, \ and\
  \bibinfo {author} {\bibfnamefont {G.}~\bibnamefont {Kotliar}},\ }\href
  {\doibase 10.1103/PhysRevLett.111.176404} {\bibfield  {journal} {\bibinfo
  {journal} {Phys.\ Rev.\ Lett.}\ }\textbf {\bibinfo {volume} {111}},\ \bibinfo
  {pages} {176404} (\bibinfo {year} {2013})}\BibitemShut {NoStop}%
\bibitem [{\citenamefont {Weng}\ \emph {et~al.}(2014)\citenamefont {Weng},
  \citenamefont {Zhao}, \citenamefont {Wang}, \citenamefont {Fang},\ and\
  \citenamefont {Dai}}]{weng_correlated_2013}%
  \BibitemOpen
  \bibfield  {author} {\bibinfo {author} {\bibfnamefont {H.}~\bibnamefont
  {Weng}}, \bibinfo {author} {\bibfnamefont {J.}~\bibnamefont {Zhao}}, \bibinfo
  {author} {\bibfnamefont {Z.}~\bibnamefont {Wang}}, \bibinfo {author}
  {\bibfnamefont {Z.}~\bibnamefont {Fang}}, \ and\ \bibinfo {author}
  {\bibfnamefont {X.}~\bibnamefont {Dai}},\ }\href {\doibase
  10.1103/PhysRevLett.112.016403} {\bibfield  {journal} {\bibinfo  {journal}
  {Phys.\ Rev.\ Lett.}\ }\textbf {\bibinfo {volume} {112}},\ \bibinfo {pages}
  {016403} (\bibinfo {year} {2014})}\BibitemShut {NoStop}%
\bibitem [{\citenamefont {Wolgast}\ \emph {et~al.}(2013)\citenamefont
  {Wolgast}, \citenamefont {Kurdak}, \citenamefont {Sun}, \citenamefont
  {Allen}, \citenamefont {Kim},\ and\ \citenamefont {Fisk}}]{wolgast_low_2013}%
  \BibitemOpen
  \bibfield  {author} {\bibinfo {author} {\bibfnamefont {S.}~\bibnamefont
  {Wolgast}}, \bibinfo {author} {\bibfnamefont {C.}~\bibnamefont {Kurdak}},
  \bibinfo {author} {\bibfnamefont {K.}~\bibnamefont {Sun}}, \bibinfo {author}
  {\bibfnamefont {J.~W.}\ \bibnamefont {Allen}}, \bibinfo {author}
  {\bibfnamefont {D.-J.}\ \bibnamefont {Kim}}, \ and\ \bibinfo {author}
  {\bibfnamefont {Z.}~\bibnamefont {Fisk}},\ }\href {\doibase
  10.1103/PhysRevB.88.180405} {\bibfield  {journal} {\bibinfo  {journal} {Phys.
  Rev. B}\ }\textbf {\bibinfo {volume} {88}},\ \bibinfo {pages} {180405}
  (\bibinfo {year} {2013})}\BibitemShut {NoStop}%
\bibitem [{\citenamefont {Kim}\ \emph {et~al.}(2013)\citenamefont {Kim},
  \citenamefont {Thomas}, \citenamefont {Grant}, \citenamefont {Botimer},
  \citenamefont {Fisk},\ and\ \citenamefont {Xia}}]{kim_robust_2013}%
  \BibitemOpen
  \bibfield  {author} {\bibinfo {author} {\bibfnamefont {D.~J.}\ \bibnamefont
  {Kim}}, \bibinfo {author} {\bibfnamefont {S.}~\bibnamefont {Thomas}},
  \bibinfo {author} {\bibfnamefont {T.}~\bibnamefont {Grant}}, \bibinfo
  {author} {\bibfnamefont {J.}~\bibnamefont {Botimer}}, \bibinfo {author}
  {\bibfnamefont {Z.}~\bibnamefont {Fisk}}, \ and\ \bibinfo {author}
  {\bibfnamefont {J.}~\bibnamefont {Xia}},\ }\href
  {http://dx.doi.org/10.1038/srep03150} {\bibfield  {journal} {\bibinfo
  {journal} {Sci.\ Rep.}\ }\textbf {\bibinfo {volume} {3}},\ \bibinfo {pages}
  {3150} (\bibinfo {year} {2013})}\BibitemShut {NoStop}%
\bibitem [{\citenamefont {Zhang}\ \emph {et~al.}(2013)\citenamefont {Zhang},
  \citenamefont {Butch}, \citenamefont {Syers}, \citenamefont {Ziemak},
  \citenamefont {Greene},\ and\ \citenamefont
  {Paglione}}]{zhang_hybridization_2013}%
  \BibitemOpen
  \bibfield  {author} {\bibinfo {author} {\bibfnamefont {X.}~\bibnamefont
  {Zhang}}, \bibinfo {author} {\bibfnamefont {N.~P.}\ \bibnamefont {Butch}},
  \bibinfo {author} {\bibfnamefont {P.}~\bibnamefont {Syers}}, \bibinfo
  {author} {\bibfnamefont {S.}~\bibnamefont {Ziemak}}, \bibinfo {author}
  {\bibfnamefont {R.~L.}\ \bibnamefont {Greene}}, \ and\ \bibinfo {author}
  {\bibfnamefont {J.}~\bibnamefont {Paglione}},\ }\href {\doibase
  10.1103/PhysRevX.3.011011} {\bibfield  {journal} {\bibinfo  {journal} {Phys.\
  Rev.\ X}\ }\textbf {\bibinfo {volume} {3}},\ \bibinfo {pages} {011011}
  (\bibinfo {year} {2013})}\BibitemShut {NoStop}%
\bibitem [{\citenamefont {Miyazaki}\ \emph {et~al.}(2012)\citenamefont
  {Miyazaki}, \citenamefont {Hajiri}, \citenamefont {Ito}, \citenamefont
  {Kunii},\ and\ \citenamefont {Kimura}}]{miyazaki_momentum_2012}%
  \BibitemOpen
  \bibfield  {author} {\bibinfo {author} {\bibfnamefont {H.}~\bibnamefont
  {Miyazaki}}, \bibinfo {author} {\bibfnamefont {T.}~\bibnamefont {Hajiri}},
  \bibinfo {author} {\bibfnamefont {T.}~\bibnamefont {Ito}}, \bibinfo {author}
  {\bibfnamefont {S.}~\bibnamefont {Kunii}}, \ and\ \bibinfo {author}
  {\bibfnamefont {S.~I.}\ \bibnamefont {Kimura}},\ }\href {\doibase
  10.1103/PhysRevB.86.075105} {\bibfield  {journal} {\bibinfo  {journal} {Phys.
  Rev. B}\ }\textbf {\bibinfo {volume} {86}},\ \bibinfo {pages} {075105}
  (\bibinfo {year} {2012})}\BibitemShut {NoStop}%
\bibitem [{\citenamefont {Xu}\ \emph {et~al.}(2013)\citenamefont {Xu},
  \citenamefont {Shi}, \citenamefont {Biswas}, \citenamefont {Matt},
  \citenamefont {Dhaka}, \citenamefont {Huang}, \citenamefont {Plumb},
  \citenamefont {Radovi\ifmmode~\acute{c}\else \'{c}\fi{}}, \citenamefont
  {Dil}, \citenamefont {Pomjakushina}, \citenamefont {Conder}, \citenamefont
  {Amato}, \citenamefont {Salman}, \citenamefont {Paul}, \citenamefont {Mesot},
  \citenamefont {Ding},\ and\ \citenamefont {Shi}}]{xu_surface_2013}%
  \BibitemOpen
  \bibfield  {author} {\bibinfo {author} {\bibfnamefont {N.}~\bibnamefont
  {Xu}}, \bibinfo {author} {\bibfnamefont {X.}~\bibnamefont {Shi}}, \bibinfo
  {author} {\bibfnamefont {P.~K.}\ \bibnamefont {Biswas}}, \bibinfo {author}
  {\bibfnamefont {C.~E.}\ \bibnamefont {Matt}}, \bibinfo {author}
  {\bibfnamefont {R.~S.}\ \bibnamefont {Dhaka}}, \bibinfo {author}
  {\bibfnamefont {Y.}~\bibnamefont {Huang}}, \bibinfo {author} {\bibfnamefont
  {N.~C.}\ \bibnamefont {Plumb}}, \bibinfo {author} {\bibfnamefont
  {M.}~\bibnamefont {Radovi\ifmmode~\acute{c}\else \'{c}\fi{}}}, \bibinfo
  {author} {\bibfnamefont {J.~H.}\ \bibnamefont {Dil}}, \bibinfo {author}
  {\bibfnamefont {E.}~\bibnamefont {Pomjakushina}}, \bibinfo {author}
  {\bibfnamefont {K.}~\bibnamefont {Conder}}, \bibinfo {author} {\bibfnamefont
  {A.}~\bibnamefont {Amato}}, \bibinfo {author} {\bibfnamefont
  {Z.}~\bibnamefont {Salman}}, \bibinfo {author} {\bibfnamefont {D.~M.}\
  \bibnamefont {Paul}}, \bibinfo {author} {\bibfnamefont {J.}~\bibnamefont
  {Mesot}}, \bibinfo {author} {\bibfnamefont {H.}~\bibnamefont {Ding}}, \ and\
  \bibinfo {author} {\bibfnamefont {M.}~\bibnamefont {Shi}},\ }\href {\doibase
  10.1103/PhysRevB.88.121102} {\bibfield  {journal} {\bibinfo  {journal}
  {Phys.\ Rev.\ B}\ }\textbf {\bibinfo {volume} {88}},\ \bibinfo {pages}
  {121102} (\bibinfo {year} {2013})}\BibitemShut {NoStop}%
\bibitem [{\citenamefont {Neupane}\ \emph {et~al.}(2013)\citenamefont
  {Neupane}, \citenamefont {Alidoust}, \citenamefont {Xu}, \citenamefont
  {Kondo}, \citenamefont {Ishida}, \citenamefont {Kim}, \citenamefont {Liu},
  \citenamefont {Belopolski}, \citenamefont {Jo}, \citenamefont {Chang},\ and\
  \citenamefont {et~al.}}]{neupane_surface_2013}%
  \BibitemOpen
  \bibfield  {author} {\bibinfo {author} {\bibfnamefont {M.}~\bibnamefont
  {Neupane}}, \bibinfo {author} {\bibfnamefont {N.}~\bibnamefont {Alidoust}},
  \bibinfo {author} {\bibfnamefont {S.-Y.}\ \bibnamefont {Xu}}, \bibinfo
  {author} {\bibfnamefont {T.}~\bibnamefont {Kondo}}, \bibinfo {author}
  {\bibfnamefont {Y.}~\bibnamefont {Ishida}}, \bibinfo {author} {\bibfnamefont
  {D.~J.}\ \bibnamefont {Kim}}, \bibinfo {author} {\bibfnamefont
  {C.}~\bibnamefont {Liu}}, \bibinfo {author} {\bibfnamefont {I.}~\bibnamefont
  {Belopolski}}, \bibinfo {author} {\bibfnamefont {Y.~J.}\ \bibnamefont {Jo}},
  \bibinfo {author} {\bibfnamefont {T.-R.}\ \bibnamefont {Chang}}, \ and\
  \bibinfo {author} {\bibnamefont {et~al.}},\ }\href
  {http://dx.doi.org/10.1038/ncomms3991} {\bibfield  {journal} {\bibinfo
  {journal} {Nat.\ Commun.}\ }\textbf {\bibinfo {volume} {4}},\ \bibinfo
  {pages} {2991} (\bibinfo {year} {2013})}\BibitemShut {NoStop}%
\bibitem [{\citenamefont {{Jiang}}\ \emph {et~al.}(2013)\citenamefont
  {{Jiang}}, \citenamefont {{Li}}, \citenamefont {{Zhang}}, \citenamefont
  {{Sun}}, \citenamefont {{Chen}}, \citenamefont {{Ye}}, \citenamefont {{Xu}},
  \citenamefont {{Ge}}, \citenamefont {{Tan}}, \citenamefont {{Niu}},
  \citenamefont {{Xia}}, \citenamefont {{Xie}}, \citenamefont {{Li}},
  \citenamefont {{Chen}}, \citenamefont {{Wen}},\ and\ \citenamefont
  {{Feng}}}]{jiang_observation_2013}%
  \BibitemOpen
  \bibfield  {author} {\bibinfo {author} {\bibfnamefont {J.}~\bibnamefont
  {{Jiang}}}, \bibinfo {author} {\bibfnamefont {S.}~\bibnamefont {{Li}}},
  \bibinfo {author} {\bibfnamefont {T.}~\bibnamefont {{Zhang}}}, \bibinfo
  {author} {\bibfnamefont {Z.}~\bibnamefont {{Sun}}}, \bibinfo {author}
  {\bibfnamefont {F.}~\bibnamefont {{Chen}}}, \bibinfo {author} {\bibfnamefont
  {Z.~R.}\ \bibnamefont {{Ye}}}, \bibinfo {author} {\bibfnamefont
  {M.}~\bibnamefont {{Xu}}}, \bibinfo {author} {\bibfnamefont {Q.~Q.}\
  \bibnamefont {{Ge}}}, \bibinfo {author} {\bibfnamefont {S.~Y.}\ \bibnamefont
  {{Tan}}}, \bibinfo {author} {\bibfnamefont {X.~H.}\ \bibnamefont {{Niu}}},
  \bibinfo {author} {\bibfnamefont {M.}~\bibnamefont {{Xia}}}, \bibinfo
  {author} {\bibfnamefont {B.~P.}\ \bibnamefont {{Xie}}}, \bibinfo {author}
  {\bibfnamefont {Y.~F.}\ \bibnamefont {{Li}}}, \bibinfo {author}
  {\bibfnamefont {X.~H.}\ \bibnamefont {{Chen}}}, \bibinfo {author}
  {\bibfnamefont {H.~H.}\ \bibnamefont {{Wen}}}, \ and\ \bibinfo {author}
  {\bibfnamefont {D.~L.}\ \bibnamefont {{Feng}}},\ }\href
  {http://dx.doi.org/10.1038/ncomms4010} {\bibfield  {journal} {\bibinfo
  {journal} {Nat.\ Commun.}\ }\textbf {\bibinfo {volume} {4}},\ \bibinfo {eid}
  {3010} (\bibinfo {year} {2013})}\BibitemShut {NoStop}%
\bibitem [{\citenamefont {Frantzeskakis}\ \emph {et~al.}(2013)\citenamefont
  {Frantzeskakis}, \citenamefont {de~Jong}, \citenamefont {Zwartsenberg},
  \citenamefont {Huang}, \citenamefont {Pan}, \citenamefont {Zhang},
  \citenamefont {Zhang}, \citenamefont {Zhang}, \citenamefont {Bao},
  \citenamefont {Tegus}, \citenamefont {Varykhalov}, \citenamefont
  {de~Visser},\ and\ \citenamefont {Golden}}]{frantzeskakis_kondo_2013}%
  \BibitemOpen
  \bibfield  {author} {\bibinfo {author} {\bibfnamefont {E.}~\bibnamefont
  {Frantzeskakis}}, \bibinfo {author} {\bibfnamefont {N.}~\bibnamefont
  {de~Jong}}, \bibinfo {author} {\bibfnamefont {B.}~\bibnamefont
  {Zwartsenberg}}, \bibinfo {author} {\bibfnamefont {Y.~K.}\ \bibnamefont
  {Huang}}, \bibinfo {author} {\bibfnamefont {Y.}~\bibnamefont {Pan}}, \bibinfo
  {author} {\bibfnamefont {X.}~\bibnamefont {Zhang}}, \bibinfo {author}
  {\bibfnamefont {J.~X.}\ \bibnamefont {Zhang}}, \bibinfo {author}
  {\bibfnamefont {F.~X.}\ \bibnamefont {Zhang}}, \bibinfo {author}
  {\bibfnamefont {L.~H.}\ \bibnamefont {Bao}}, \bibinfo {author} {\bibfnamefont
  {O.}~\bibnamefont {Tegus}}, \bibinfo {author} {\bibfnamefont
  {A.}~\bibnamefont {Varykhalov}}, \bibinfo {author} {\bibfnamefont
  {A.}~\bibnamefont {de~Visser}}, \ and\ \bibinfo {author} {\bibfnamefont
  {M.~S.}\ \bibnamefont {Golden}},\ }\href {\doibase 10.1103/PhysRevX.3.041024}
  {\bibfield  {journal} {\bibinfo  {journal} {Phys. Rev. X}\ }\textbf {\bibinfo
  {volume} {3}},\ \bibinfo {pages} {041024} (\bibinfo {year}
  {2013})}\BibitemShut {NoStop}%
\bibitem [{\citenamefont {Min}\ \emph {et~al.}(2014)\citenamefont {Min},
  \citenamefont {Lutz}, \citenamefont {Fiedler}, \citenamefont {Kang},
  \citenamefont {Cho}, \citenamefont {Kim}, \citenamefont {Bentmann},\ and\
  \citenamefont {Reinert}}]{min_importance_2013}%
  \BibitemOpen
  \bibfield  {author} {\bibinfo {author} {\bibfnamefont {C.-H.}\ \bibnamefont
  {Min}}, \bibinfo {author} {\bibfnamefont {P.}~\bibnamefont {Lutz}}, \bibinfo
  {author} {\bibfnamefont {S.}~\bibnamefont {Fiedler}}, \bibinfo {author}
  {\bibfnamefont {B.~Y.}\ \bibnamefont {Kang}}, \bibinfo {author}
  {\bibfnamefont {B.~K.}\ \bibnamefont {Cho}}, \bibinfo {author} {\bibfnamefont
  {H.-D.}\ \bibnamefont {Kim}}, \bibinfo {author} {\bibfnamefont
  {H.}~\bibnamefont {Bentmann}}, \ and\ \bibinfo {author} {\bibfnamefont
  {F.}~\bibnamefont {Reinert}},\ }\href {\doibase
  10.1103/PhysRevLett.112.226402} {\bibfield  {journal} {\bibinfo  {journal}
  {Phys. Rev. Lett.}\ }\textbf {\bibinfo {volume} {112}},\ \bibinfo {pages}
  {226402} (\bibinfo {year} {2014})}\BibitemShut {NoStop}%
\bibitem [{\citenamefont {Li}\ \emph {et~al.}(2014)\citenamefont {Li},
  \citenamefont {Xiang}, \citenamefont {Yu}, \citenamefont {Asaba},
  \citenamefont {Lawson}, \citenamefont {Cai}, \citenamefont {Tinsman},
  \citenamefont {Berkley}, \citenamefont {Wolgast}, \citenamefont {Eo},
  \citenamefont {Kim}, \citenamefont {Kurdak}, \citenamefont {Allen},
  \citenamefont {Sun}, \citenamefont {Chen}, \citenamefont {Wang},
  \citenamefont {Fisk},\ and\ \citenamefont {Li}}]{li_quantum_2013}%
  \BibitemOpen
  \bibfield  {author} {\bibinfo {author} {\bibfnamefont {G.}~\bibnamefont
  {Li}}, \bibinfo {author} {\bibfnamefont {Z.}~\bibnamefont {Xiang}}, \bibinfo
  {author} {\bibfnamefont {F.}~\bibnamefont {Yu}}, \bibinfo {author}
  {\bibfnamefont {T.}~\bibnamefont {Asaba}}, \bibinfo {author} {\bibfnamefont
  {B.}~\bibnamefont {Lawson}}, \bibinfo {author} {\bibfnamefont
  {P.}~\bibnamefont {Cai}}, \bibinfo {author} {\bibfnamefont {C.}~\bibnamefont
  {Tinsman}}, \bibinfo {author} {\bibfnamefont {A.}~\bibnamefont {Berkley}},
  \bibinfo {author} {\bibfnamefont {S.}~\bibnamefont {Wolgast}}, \bibinfo
  {author} {\bibfnamefont {Y.~S.}\ \bibnamefont {Eo}}, \bibinfo {author}
  {\bibfnamefont {D.-J.}\ \bibnamefont {Kim}}, \bibinfo {author} {\bibfnamefont
  {C.}~\bibnamefont {Kurdak}}, \bibinfo {author} {\bibfnamefont {J.~W.}\
  \bibnamefont {Allen}}, \bibinfo {author} {\bibfnamefont {K.}~\bibnamefont
  {Sun}}, \bibinfo {author} {\bibfnamefont {X.~H.}\ \bibnamefont {Chen}},
  \bibinfo {author} {\bibfnamefont {Y.~Y.}\ \bibnamefont {Wang}}, \bibinfo
  {author} {\bibfnamefont {Z.}~\bibnamefont {Fisk}}, \ and\ \bibinfo {author}
  {\bibfnamefont {L.}~\bibnamefont {Li}},\ }\href {\doibase
  10.1126/science.1250366} {\bibfield  {journal} {\bibinfo  {journal}
  {Science}\ }\textbf {\bibinfo {volume} {346}},\ \bibinfo {pages} {1208}
  (\bibinfo {year} {2014})},\ \Eprint
  {http://arxiv.org/abs/http://www.sciencemag.org/content/346/6214/1208.full.pdf}
  {http://www.sciencemag.org/content/346/6214/1208.full.pdf} \BibitemShut
  {NoStop}%
\bibitem [{\citenamefont {{Yee}}\ \emph {et~al.}(2013)\citenamefont {{Yee}},
  \citenamefont {{He}}, \citenamefont {{Soumyanarayanan}}, \citenamefont
  {{Kim}}, \citenamefont {{Fisk}},\ and\ \citenamefont
  {{Hoffman}}}]{yee_imaging_2013}%
  \BibitemOpen
  \bibfield  {author} {\bibinfo {author} {\bibfnamefont {M.~M.}\ \bibnamefont
  {{Yee}}}, \bibinfo {author} {\bibfnamefont {Y.}~\bibnamefont {{He}}},
  \bibinfo {author} {\bibfnamefont {A.}~\bibnamefont {{Soumyanarayanan}}},
  \bibinfo {author} {\bibfnamefont {D.-J.}\ \bibnamefont {{Kim}}}, \bibinfo
  {author} {\bibfnamefont {Z.}~\bibnamefont {{Fisk}}}, \ and\ \bibinfo {author}
  {\bibfnamefont {J.~E.}\ \bibnamefont {{Hoffman}}},\ }\href@noop {} {\bibfield
   {journal} {\bibinfo  {journal} {ArXiv e-prints}\ } (\bibinfo {year}
  {2013})},\ \Eprint {http://arxiv.org/abs/1308.1085} {arXiv:1308.1085
  [cond-mat.str-el]} \BibitemShut {NoStop}%
\bibitem [{\citenamefont {Gorshunov}\ \emph {et~al.}(1999)\citenamefont
  {Gorshunov}, \citenamefont {Sluchanko}, \citenamefont {Volkov}, \citenamefont
  {Dressel}, \citenamefont {Knebel}, \citenamefont {Loidl},\ and\ \citenamefont
  {Kunii}}]{gorshunov_smb6gap_1999}%
  \BibitemOpen
  \bibfield  {author} {\bibinfo {author} {\bibfnamefont {B.}~\bibnamefont
  {Gorshunov}}, \bibinfo {author} {\bibfnamefont {N.}~\bibnamefont
  {Sluchanko}}, \bibinfo {author} {\bibfnamefont {A.}~\bibnamefont {Volkov}},
  \bibinfo {author} {\bibfnamefont {M.}~\bibnamefont {Dressel}}, \bibinfo
  {author} {\bibfnamefont {G.}~\bibnamefont {Knebel}}, \bibinfo {author}
  {\bibfnamefont {A.}~\bibnamefont {Loidl}}, \ and\ \bibinfo {author}
  {\bibfnamefont {S.}~\bibnamefont {Kunii}},\ }\href {\doibase
  10.1103/PhysRevB.59.1808} {\bibfield  {journal} {\bibinfo  {journal} {Phys.
  Rev. B}\ }\textbf {\bibinfo {volume} {59}},\ \bibinfo {pages} {1808}
  (\bibinfo {year} {1999})}\BibitemShut {NoStop}%
\bibitem [{\citenamefont {Xu}\ \emph {et~al.}(2014)\citenamefont {Xu},
  \citenamefont {Biswas}, \citenamefont {Dil}, \citenamefont {Dhaka},
  \citenamefont {Landolt}, \citenamefont {Muff}, \citenamefont {Matt},
  \citenamefont {Shi}, \citenamefont {Plumb}, \citenamefont {Radovi{\'{c}}},
  \citenamefont {Pomjakushina}, \citenamefont {Conder}, \citenamefont {Amato},
  \citenamefont {Borisenko}, \citenamefont {Yu}, \citenamefont {Weng},
  \citenamefont {Fang}, \citenamefont {Dai}, \citenamefont {Mesot},
  \citenamefont {Ding},\ and\ \citenamefont {Shi}}]{xu_spin_2014}%
  \BibitemOpen
  \bibfield  {author} {\bibinfo {author} {\bibfnamefont {N.}~\bibnamefont
  {Xu}}, \bibinfo {author} {\bibfnamefont {P.~K.}\ \bibnamefont {Biswas}},
  \bibinfo {author} {\bibfnamefont {J.~H.}\ \bibnamefont {Dil}}, \bibinfo
  {author} {\bibfnamefont {R.~S.}\ \bibnamefont {Dhaka}}, \bibinfo {author}
  {\bibfnamefont {G.}~\bibnamefont {Landolt}}, \bibinfo {author} {\bibfnamefont
  {S.}~\bibnamefont {Muff}}, \bibinfo {author} {\bibfnamefont {C.~E.}\
  \bibnamefont {Matt}}, \bibinfo {author} {\bibfnamefont {X.}~\bibnamefont
  {Shi}}, \bibinfo {author} {\bibfnamefont {N.~C.}\ \bibnamefont {Plumb}},
  \bibinfo {author} {\bibfnamefont {M.}~\bibnamefont {Radovi{\'{c}}}}, \bibinfo
  {author} {\bibfnamefont {E.}~\bibnamefont {Pomjakushina}}, \bibinfo {author}
  {\bibfnamefont {K.}~\bibnamefont {Conder}}, \bibinfo {author} {\bibfnamefont
  {A.}~\bibnamefont {Amato}}, \bibinfo {author} {\bibfnamefont {S.~V.}\
  \bibnamefont {Borisenko}}, \bibinfo {author} {\bibfnamefont {R.}~\bibnamefont
  {Yu}}, \bibinfo {author} {\bibfnamefont {H.-M.}\ \bibnamefont {Weng}},
  \bibinfo {author} {\bibfnamefont {Z.}~\bibnamefont {Fang}}, \bibinfo {author}
  {\bibfnamefont {X.}~\bibnamefont {Dai}}, \bibinfo {author} {\bibfnamefont
  {J.}~\bibnamefont {Mesot}}, \bibinfo {author} {\bibfnamefont
  {H.}~\bibnamefont {Ding}}, \ and\ \bibinfo {author} {\bibfnamefont
  {M.}~\bibnamefont {Shi}},\ }\href {http://dx.doi.org/10.1038/ncomms5566}
  {\bibfield  {journal} {\bibinfo  {journal} {Nat Comms}\ }\textbf {\bibinfo
  {volume} {5}},\ \bibinfo {pages} {4566} (\bibinfo {year} {2014})}\BibitemShut
  {NoStop}%
\bibitem [{\citenamefont {{Hlawenka}}\ \emph {et~al.}(2015)\citenamefont
  {{Hlawenka}}, \citenamefont {{Siemensmeyer}}, \citenamefont {{Weschke}},
  \citenamefont {{Varykhalov}}, \citenamefont {{S{\'a}nchez-Barriga}},
  \citenamefont {{Shitsevalova}}, \citenamefont {{Dukhnenko}}, \citenamefont
  {{Filipov}}, \citenamefont {{Gab{\'a}ni}}, \citenamefont {{Flachbart}},
  \citenamefont {{Rader}},\ and\ \citenamefont
  {{Rienks}}}]{hlawenka_trivial_2015}%
  \BibitemOpen
  \bibfield  {author} {\bibinfo {author} {\bibfnamefont {P.}~\bibnamefont
  {{Hlawenka}}}, \bibinfo {author} {\bibfnamefont {K.}~\bibnamefont
  {{Siemensmeyer}}}, \bibinfo {author} {\bibfnamefont {E.}~\bibnamefont
  {{Weschke}}}, \bibinfo {author} {\bibfnamefont {A.}~\bibnamefont
  {{Varykhalov}}}, \bibinfo {author} {\bibfnamefont {J.}~\bibnamefont
  {{S{\'a}nchez-Barriga}}}, \bibinfo {author} {\bibfnamefont {N.~Y.}\
  \bibnamefont {{Shitsevalova}}}, \bibinfo {author} {\bibfnamefont {A.~V.}\
  \bibnamefont {{Dukhnenko}}}, \bibinfo {author} {\bibfnamefont {V.~B.}\
  \bibnamefont {{Filipov}}}, \bibinfo {author} {\bibfnamefont {S.}~\bibnamefont
  {{Gab{\'a}ni}}}, \bibinfo {author} {\bibfnamefont {K.}~\bibnamefont
  {{Flachbart}}}, \bibinfo {author} {\bibfnamefont {O.}~\bibnamefont
  {{Rader}}}, \ and\ \bibinfo {author} {\bibfnamefont {E.~D.~L.}\ \bibnamefont
  {{Rienks}}},\ }\href@noop {} {\bibfield  {journal} {\bibinfo  {journal}
  {ArXiv e-prints}\ } (\bibinfo {year} {2015})},\ \Eprint
  {http://arxiv.org/abs/1502.01542} {arXiv:1502.01542 [cond-mat.str-el]}
  \BibitemShut {NoStop}%
\bibitem [{\citenamefont {Heming}\ \emph {et~al.}(2014)\citenamefont {Heming},
  \citenamefont {Treske}, \citenamefont {Knupfer}, \citenamefont {B\"uchner},
  \citenamefont {Inosov}, \citenamefont {Shitsevalova}, \citenamefont
  {Filipov}, \citenamefont {Krause},\ and\ \citenamefont
  {Koitzsch}}]{heming_surface_2014}%
  \BibitemOpen
  \bibfield  {author} {\bibinfo {author} {\bibfnamefont {N.}~\bibnamefont
  {Heming}}, \bibinfo {author} {\bibfnamefont {U.}~\bibnamefont {Treske}},
  \bibinfo {author} {\bibfnamefont {M.}~\bibnamefont {Knupfer}}, \bibinfo
  {author} {\bibfnamefont {B.}~\bibnamefont {B\"uchner}}, \bibinfo {author}
  {\bibfnamefont {D.~S.}\ \bibnamefont {Inosov}}, \bibinfo {author}
  {\bibfnamefont {N.~Y.}\ \bibnamefont {Shitsevalova}}, \bibinfo {author}
  {\bibfnamefont {V.~B.}\ \bibnamefont {Filipov}}, \bibinfo {author}
  {\bibfnamefont {S.}~\bibnamefont {Krause}}, \ and\ \bibinfo {author}
  {\bibfnamefont {A.}~\bibnamefont {Koitzsch}},\ }\href {\doibase
  10.1103/PhysRevB.90.195128} {\bibfield  {journal} {\bibinfo  {journal} {Phys.
  Rev. B}\ }\textbf {\bibinfo {volume} {90}},\ \bibinfo {pages} {195128}
  (\bibinfo {year} {2014})}\BibitemShut {NoStop}%
\bibitem [{\citenamefont {Alexandrov}\ \emph {et~al.}(2015)\citenamefont
  {Alexandrov}, \citenamefont {Coleman},\ and\ \citenamefont
  {Erten}}]{alexandrov_breakdown_2015}%
  \BibitemOpen
  \bibfield  {author} {\bibinfo {author} {\bibfnamefont {V.}~\bibnamefont
  {Alexandrov}}, \bibinfo {author} {\bibfnamefont {P.}~\bibnamefont {Coleman}},
  \ and\ \bibinfo {author} {\bibfnamefont {O.}~\bibnamefont {Erten}},\ }\href
  {\doibase 10.1103/PhysRevLett.114.177202} {\bibfield  {journal} {\bibinfo
  {journal} {Phys. Rev. Lett.}\ }\textbf {\bibinfo {volume} {114}},\ \bibinfo
  {pages} {177202} (\bibinfo {year} {2015})}\BibitemShut {NoStop}%
\bibitem [{\citenamefont {Kapilevich}\ \emph {et~al.}(2015)\citenamefont
  {Kapilevich}, \citenamefont {Riseborough}, \citenamefont {Gray},
  \citenamefont {Gulacsi}, \citenamefont {Durakiewicz},\ and\ \citenamefont
  {Smith}}]{kapilevich_incomplete_2015}%
  \BibitemOpen
  \bibfield  {author} {\bibinfo {author} {\bibfnamefont {G.~A.}\ \bibnamefont
  {Kapilevich}}, \bibinfo {author} {\bibfnamefont {P.~S.}\ \bibnamefont
  {Riseborough}}, \bibinfo {author} {\bibfnamefont {A.~X.}\ \bibnamefont
  {Gray}}, \bibinfo {author} {\bibfnamefont {M.}~\bibnamefont {Gulacsi}},
  \bibinfo {author} {\bibfnamefont {T.}~\bibnamefont {Durakiewicz}}, \ and\
  \bibinfo {author} {\bibfnamefont {J.~L.}\ \bibnamefont {Smith}},\ }\href
  {\doibase 10.1103/PhysRevB.92.085133} {\bibfield  {journal} {\bibinfo
  {journal} {Phys. Rev. B}\ }\textbf {\bibinfo {volume} {92}},\ \bibinfo
  {pages} {085133} (\bibinfo {year} {2015})}\BibitemShut {NoStop}%
\bibitem [{\citenamefont {Zhu}\ \emph {et~al.}(2013)\citenamefont {Zhu},
  \citenamefont {Nicolaou}, \citenamefont {Levy}, \citenamefont {Butch},
  \citenamefont {Syers}, \citenamefont {Wang}, \citenamefont {Paglione},
  \citenamefont {Sawatzky}, \citenamefont {Elfimov},\ and\ \citenamefont
  {Damascelli}}]{zhu_polarity_2013}%
  \BibitemOpen
  \bibfield  {author} {\bibinfo {author} {\bibfnamefont {Z.-H.}\ \bibnamefont
  {Zhu}}, \bibinfo {author} {\bibfnamefont {A.}~\bibnamefont {Nicolaou}},
  \bibinfo {author} {\bibfnamefont {G.}~\bibnamefont {Levy}}, \bibinfo {author}
  {\bibfnamefont {N.~P.}\ \bibnamefont {Butch}}, \bibinfo {author}
  {\bibfnamefont {P.}~\bibnamefont {Syers}}, \bibinfo {author} {\bibfnamefont
  {X.~F.}\ \bibnamefont {Wang}}, \bibinfo {author} {\bibfnamefont
  {J.}~\bibnamefont {Paglione}}, \bibinfo {author} {\bibfnamefont {G.~A.}\
  \bibnamefont {Sawatzky}}, \bibinfo {author} {\bibfnamefont {I.~S.}\
  \bibnamefont {Elfimov}}, \ and\ \bibinfo {author} {\bibfnamefont
  {A.}~\bibnamefont {Damascelli}},\ }\href {\doibase
  10.1103/PhysRevLett.111.216402} {\bibfield  {journal} {\bibinfo  {journal}
  {Phys. Rev. Lett.}\ }\textbf {\bibinfo {volume} {111}},\ \bibinfo {pages}
  {216402} (\bibinfo {year} {2013})}\BibitemShut {NoStop}%
\bibitem [{\citenamefont {Legner}\ \emph {et~al.}()\citenamefont {Legner},
  \citenamefont {R\"uegg},\ and\ \citenamefont {Sigrist}}]{sm}%
  \BibitemOpen
  \bibfield  {author} {\bibinfo {author} {\bibfnamefont {M.}~\bibnamefont
  {Legner}}, \bibinfo {author} {\bibfnamefont {A.}~\bibnamefont {R\"uegg}}, \
  and\ \bibinfo {author} {\bibfnamefont {M.}~\bibnamefont {Sigrist}},\
  }\href@noop {} {\emph {\bibinfo {title} {Supplemental Material}}}\BibitemShut
  {NoStop}%
\bibitem [{\citenamefont {{Yu}}\ \emph {et~al.}(2015)\citenamefont {{Yu}},
  \citenamefont {{Weng}}, \citenamefont {{Hu}}, \citenamefont {{Fang}},\ and\
  \citenamefont {{Dai}}}]{yu_model_2014}%
  \BibitemOpen
  \bibfield  {author} {\bibinfo {author} {\bibfnamefont {R.}~\bibnamefont
  {{Yu}}}, \bibinfo {author} {\bibfnamefont {H.}~\bibnamefont {{Weng}}},
  \bibinfo {author} {\bibfnamefont {X.}~\bibnamefont {{Hu}}}, \bibinfo {author}
  {\bibfnamefont {Z.}~\bibnamefont {{Fang}}}, \ and\ \bibinfo {author}
  {\bibfnamefont {X.}~\bibnamefont {{Dai}}},\ }\href {\doibase
  10.1088/1367-2630/17/2/023012} {\bibfield  {journal} {\bibinfo  {journal}
  {New J.\ Phys.}\ }\textbf {\bibinfo {volume} {17}},\ \bibinfo {eid} {023012}
  (\bibinfo {year} {2015})}\BibitemShut {NoStop}%
\bibitem [{\citenamefont {Kim}\ \emph {et~al.}(2014)\citenamefont {Kim},
  \citenamefont {Kim}, \citenamefont {Kang}, \citenamefont {Kim}, \citenamefont
  {Choi}, \citenamefont {Kang}, \citenamefont {Denlinger},\ and\ \citenamefont
  {Min}}]{kim_termination_2014}%
  \BibitemOpen
  \bibfield  {author} {\bibinfo {author} {\bibfnamefont {J.}~\bibnamefont
  {Kim}}, \bibinfo {author} {\bibfnamefont {K.}~\bibnamefont {Kim}}, \bibinfo
  {author} {\bibfnamefont {C.-J.}\ \bibnamefont {Kang}}, \bibinfo {author}
  {\bibfnamefont {S.}~\bibnamefont {Kim}}, \bibinfo {author} {\bibfnamefont
  {H.~C.}\ \bibnamefont {Choi}}, \bibinfo {author} {\bibfnamefont {J.-S.}\
  \bibnamefont {Kang}}, \bibinfo {author} {\bibfnamefont {J.~D.}\ \bibnamefont
  {Denlinger}}, \ and\ \bibinfo {author} {\bibfnamefont {B.~I.}\ \bibnamefont
  {Min}},\ }\href {\doibase 10.1103/PhysRevB.90.075131} {\bibfield  {journal}
  {\bibinfo  {journal} {Phys. Rev. B}\ }\textbf {\bibinfo {volume} {90}},\
  \bibinfo {pages} {075131} (\bibinfo {year} {2014})}\BibitemShut {NoStop}%
\bibitem [{\citenamefont {Baruselli}\ and\ \citenamefont
  {Vojta}(2014)}]{baruselli_scanning_2014}%
  \BibitemOpen
  \bibfield  {author} {\bibinfo {author} {\bibfnamefont {P.~P.}\ \bibnamefont
  {Baruselli}}\ and\ \bibinfo {author} {\bibfnamefont {M.}~\bibnamefont
  {Vojta}},\ }\href {\doibase 10.1103/PhysRevB.90.201106} {\bibfield  {journal}
  {\bibinfo  {journal} {Phys. Rev. B}\ }\textbf {\bibinfo {volume} {90}},\
  \bibinfo {pages} {201106} (\bibinfo {year} {2014})}\BibitemShut {NoStop}%
\bibitem [{Note1()}]{Note1}%
  \BibitemOpen
  \bibinfo {note} {The {\protect \MakeUppercase {\relax \protect \fontsize
  {9}{10.5}\protect \selectfont \abovedisplayskip 8.5\p@ plus3\p@ minus4\p@
  \belowdisplayskip \abovedisplayskip \abovedisplayshortskip \z@ plus2\p@
  \belowdisplayshortskip 4\p@ plus2\p@ minus2\p@ \def \leftmargin \leftmargini
  \parsep 4\p@ plus2\p@ minus\p@ \topsep 8\p@ plus2\p@ minus4\p@ \itemsep 4\p@
  plus2\p@ minus\p@ {\leftmargin \leftmargini \topsep 4\p@ plus2\p@ minus2\p@
  \parsep 2\p@ plus\p@ minus\p@ \itemsep \parsep }hsp}}{} ${\protect \bm {K}}$
  has the property $-{\protect \bm {K}}={\protect \bm {K}}+{\protect \bm {G}}$,
  where ${\protect \bm {G}}$ is a reciprocal lattice vector. On the (001)
  surface, there are three different {\protect \MakeUppercase {\relax \protect
  \fontsize {9}{10.5}\protect \selectfont \abovedisplayskip 8.5\p@ plus3\p@
  minus4\p@ \belowdisplayskip \abovedisplayskip \abovedisplayshortskip \z@
  plus2\p@ \belowdisplayshortskip 4\p@ plus2\p@ minus2\p@ \def \leftmargin
  \leftmargini \parsep 4\p@ plus2\p@ minus\p@ \topsep 8\p@ plus2\p@ minus4\p@
  \itemsep 4\p@ plus2\p@ minus\p@ {\leftmargin \leftmargini \topsep 4\p@
  plus2\p@ minus2\p@ \parsep 2\p@ plus\p@ minus\p@ \itemsep \parsep }hsp}}{}s:
  $\protect \mathrm {\bar \Gamma }$, $\protect \mathrm {\bar X}$, and $\protect
  \mathrm {\bar M}$.}\BibitemShut {Stop}%
\bibitem [{\citenamefont {Fu}(2011)}]{fu_topological_2011}%
  \BibitemOpen
  \bibfield  {author} {\bibinfo {author} {\bibfnamefont {L.}~\bibnamefont
  {Fu}},\ }\href {\doibase 10.1103/PhysRevLett.106.106802} {\bibfield
  {journal} {\bibinfo  {journal} {Phys.\ Rev.\ Lett.}\ }\textbf {\bibinfo
  {volume} {106}},\ \bibinfo {pages} {106802} (\bibinfo {year}
  {2011})}\BibitemShut {NoStop}%
\bibitem [{\citenamefont {Legner}\ \emph {et~al.}(2014)\citenamefont {Legner},
  \citenamefont {R\"uegg},\ and\ \citenamefont
  {Sigrist}}]{legner_topological_2014}%
  \BibitemOpen
  \bibfield  {author} {\bibinfo {author} {\bibfnamefont {M.}~\bibnamefont
  {Legner}}, \bibinfo {author} {\bibfnamefont {A.}~\bibnamefont {R\"uegg}}, \
  and\ \bibinfo {author} {\bibfnamefont {M.}~\bibnamefont {Sigrist}},\ }\href
  {\doibase 10.1103/PhysRevB.89.085110} {\bibfield  {journal} {\bibinfo
  {journal} {Phys. Rev. B}\ }\textbf {\bibinfo {volume} {89}},\ \bibinfo
  {pages} {085110} (\bibinfo {year} {2014})}\BibitemShut {NoStop}%
\bibitem [{\citenamefont {Wang}\ \emph {et~al.}(2015)\citenamefont {Wang},
  \citenamefont {Wu}, \citenamefont {Felser}, \citenamefont {Yan},\ and\
  \citenamefont {Liu}}]{wang_spin_2014}%
  \BibitemOpen
  \bibfield  {author} {\bibinfo {author} {\bibfnamefont {Q.-Z.}\ \bibnamefont
  {Wang}}, \bibinfo {author} {\bibfnamefont {S.-C.}\ \bibnamefont {Wu}},
  \bibinfo {author} {\bibfnamefont {C.}~\bibnamefont {Felser}}, \bibinfo
  {author} {\bibfnamefont {B.}~\bibnamefont {Yan}}, \ and\ \bibinfo {author}
  {\bibfnamefont {C.-X.}\ \bibnamefont {Liu}},\ }\href {\doibase
  10.1103/PhysRevB.91.165435} {\bibfield  {journal} {\bibinfo  {journal} {Phys.
  Rev. B}\ }\textbf {\bibinfo {volume} {91}},\ \bibinfo {pages} {165435}
  (\bibinfo {year} {2015})}\BibitemShut {NoStop}%
\bibitem [{\citenamefont {Kang}\ \emph {et~al.}(2015)\citenamefont {Kang},
  \citenamefont {Kim}, \citenamefont {Kim}, \citenamefont {Kang}, \citenamefont
  {Denlinger},\ and\ \citenamefont {Min}}]{kang_band_2013}%
  \BibitemOpen
  \bibfield  {author} {\bibinfo {author} {\bibfnamefont {C.-J.}\ \bibnamefont
  {Kang}}, \bibinfo {author} {\bibfnamefont {J.}~\bibnamefont {Kim}}, \bibinfo
  {author} {\bibfnamefont {K.}~\bibnamefont {Kim}}, \bibinfo {author}
  {\bibfnamefont {J.}~\bibnamefont {Kang}}, \bibinfo {author} {\bibfnamefont
  {J.~D.}\ \bibnamefont {Denlinger}}, \ and\ \bibinfo {author} {\bibfnamefont
  {B.~I.}\ \bibnamefont {Min}},\ }\href {\doibase 10.7566/JPSJ.84.024722}
  {\bibfield  {journal} {\bibinfo  {journal} {J. Phys. Soc. Jpn.}\ }\textbf
  {\bibinfo {volume} {84}},\ \bibinfo {pages} {024722} (\bibinfo {year}
  {2015})}\BibitemShut {NoStop}%
\bibitem [{Note2()}]{Note2}%
  \BibitemOpen
  \bibinfo {note} {This orbital pseudospin should not be confused with the
  surface-state pseudospin defined \protect \hyperref [s:pseudospin]{before}.
  The detailed relation between the orbital pseudospin and the physical spin of
  the $f$ electrons is given in \protect \hyper@link
  {cite}{cite.sm}{Ref.~\protect \NoHyper \protect \rev@citealp {sm}\protect
  \endNoHyper }.}\BibitemShut {Stop}%
\bibitem [{\citenamefont {Read}\ and\ \citenamefont
  {Newns}(1983)}]{Read:1983b}%
  \BibitemOpen
  \bibfield  {author} {\bibinfo {author} {\bibfnamefont {N.}~\bibnamefont
  {Read}}\ and\ \bibinfo {author} {\bibfnamefont {D.~M.}\ \bibnamefont
  {Newns}},\ }\href {http://stacks.iop.org/0022-3719/16/L1055} {\bibfield
  {journal} {\bibinfo  {journal} {J.\ Phys.\ C}\ }\textbf {\bibinfo {volume}
  {16}},\ \bibinfo {pages} {L1055} (\bibinfo {year} {1983})}\BibitemShut
  {NoStop}%
\bibitem [{\citenamefont {Coleman}(1984)}]{Coleman:1984}%
  \BibitemOpen
  \bibfield  {author} {\bibinfo {author} {\bibfnamefont {P.}~\bibnamefont
  {Coleman}},\ }\href {\doibase 10.1103/PhysRevB.29.3035} {\bibfield  {journal}
  {\bibinfo  {journal} {Phys.\ Rev.\ B}\ }\textbf {\bibinfo {volume} {29}},\
  \bibinfo {pages} {3035} (\bibinfo {year} {1984})}\BibitemShut {NoStop}%
\bibitem [{\citenamefont {Rice}\ and\ \citenamefont {Ueda}(1985)}]{Rice:1985b}%
  \BibitemOpen
  \bibfield  {author} {\bibinfo {author} {\bibfnamefont {T.~M.}\ \bibnamefont
  {Rice}}\ and\ \bibinfo {author} {\bibfnamefont {K.}~\bibnamefont {Ueda}},\
  }\href {http://link.aps.org/abstract/PRL/v55/p995} {\bibfield  {journal}
  {\bibinfo  {journal} {Phys.\ Rev.\ Lett.}\ }\textbf {\bibinfo {volume}
  {55}},\ \bibinfo {pages} {995} (\bibinfo {year} {1985})}\BibitemShut
  {NoStop}%
\bibitem [{\citenamefont {Wang}\ and\ \citenamefont
  {Zhang}(2012)}]{wang_simplified_2012}%
  \BibitemOpen
  \bibfield  {author} {\bibinfo {author} {\bibfnamefont {Z.}~\bibnamefont
  {Wang}}\ and\ \bibinfo {author} {\bibfnamefont {S.-C.}\ \bibnamefont
  {Zhang}},\ }\href {\doibase 10.1103/PhysRevX.2.031008} {\bibfield  {journal}
  {\bibinfo  {journal} {Phys. Rev. X}\ }\textbf {\bibinfo {volume} {2}},\
  \bibinfo {pages} {031008} (\bibinfo {year} {2012})}\BibitemShut {NoStop}%
\bibitem [{\citenamefont {Hohenadler}\ and\ \citenamefont
  {Assaad}(2013)}]{hohenadler_correlation_2013}%
  \BibitemOpen
  \bibfield  {author} {\bibinfo {author} {\bibfnamefont {M.}~\bibnamefont
  {Hohenadler}}\ and\ \bibinfo {author} {\bibfnamefont {F.~F.}\ \bibnamefont
  {Assaad}},\ }\href {http://stacks.iop.org/0953-8984/25/i=14/a=143201}
  {\bibfield  {journal} {\bibinfo  {journal} {Journal of Physics: Condensed
  Matter}\ }\textbf {\bibinfo {volume} {25}},\ \bibinfo {pages} {143201}
  (\bibinfo {year} {2013})}\BibitemShut {NoStop}%
\bibitem [{\citenamefont {Werner}\ and\ \citenamefont
  {Assaad}(2013)}]{werner_interaction_2013}%
  \BibitemOpen
  \bibfield  {author} {\bibinfo {author} {\bibfnamefont {J.}~\bibnamefont
  {Werner}}\ and\ \bibinfo {author} {\bibfnamefont {F.~F.}\ \bibnamefont
  {Assaad}},\ }\href {\doibase 10.1103/PhysRevB.88.035113} {\bibfield
  {journal} {\bibinfo  {journal} {Phys.\ Rev.\ B}\ }\textbf {\bibinfo {volume}
  {88}},\ \bibinfo {pages} {035113} (\bibinfo {year} {2013})}\BibitemShut
  {NoStop}%
\bibitem [{\citenamefont {Fukui}\ \emph {et~al.}(2005)\citenamefont {Fukui},
  \citenamefont {Hatsugai},\ and\ \citenamefont {Suzuki}}]{suzuki_chern_2005}%
  \BibitemOpen
  \bibfield  {author} {\bibinfo {author} {\bibfnamefont {T.}~\bibnamefont
  {Fukui}}, \bibinfo {author} {\bibfnamefont {Y.}~\bibnamefont {Hatsugai}}, \
  and\ \bibinfo {author} {\bibfnamefont {H.}~\bibnamefont {Suzuki}},\ }\href
  {\doibase 10.1143/JPSJ.74.1674} {\bibfield  {journal} {\bibinfo  {journal}
  {J.\ Phys.\ Soc.\ Jpn.}\ }\textbf {\bibinfo {volume} {74}},\ \bibinfo {pages}
  {1674} (\bibinfo {year} {2005})}\BibitemShut {NoStop}%
\bibitem [{\citenamefont {Baruselli}\ and\ \citenamefont
  {Vojta}(2015)}]{baruselli_distinct_2015}%
  \BibitemOpen
  \bibfield  {author} {\bibinfo {author} {\bibfnamefont {P.~P.}\ \bibnamefont
  {Baruselli}}\ and\ \bibinfo {author} {\bibfnamefont {M.}~\bibnamefont
  {Vojta}},\ }\href {\doibase 10.1103/PhysRevLett.115.156404} {\bibfield
  {journal} {\bibinfo  {journal} {Phys. Rev. Lett.}\ }\textbf {\bibinfo
  {volume} {115}},\ \bibinfo {pages} {156404} (\bibinfo {year}
  {2015})}\BibitemShut {NoStop}%
\end{thebibliography}

\begin{thebibliography}{7}%
\makeatletter
\providecommand \@ifxundefined [1]{%
 \@ifx{#1\undefined}
}%
\providecommand \@ifnum [1]{%
 \ifnum #1\expandafter \@firstoftwo
 \else \expandafter \@secondoftwo
 \fi
}%
\providecommand \@ifx [1]{%
 \ifx #1\expandafter \@firstoftwo
 \else \expandafter \@secondoftwo
 \fi
}%
\providecommand \natexlab [1]{#1}%
\providecommand \enquote  [1]{``#1''}%
\providecommand \bibnamefont  [1]{#1}%
\providecommand \bibfnamefont [1]{#1}%
\providecommand \citenamefont [1]{#1}%
\providecommand \href@noop [0]{\@secondoftwo}%
\providecommand \href [0]{\begingroup \@sanitize@url \@href}%
\providecommand \@href[1]{\@@startlink{#1}\@@href}%
\providecommand \@@href[1]{\endgroup#1\@@endlink}%
\providecommand \@sanitize@url [0]{\catcode `\\12\catcode `\$12\catcode
  `\&12\catcode `\#12\catcode `\^12\catcode `\_12\catcode `\%12\relax}%
\providecommand \@@startlink[1]{}%
\providecommand \@@endlink[0]{}%
\providecommand \url  [0]{\begingroup\@sanitize@url \@url }%
\providecommand \@url [1]{\endgroup\@href {#1}{\urlprefix }}%
\providecommand \urlprefix  [0]{URL }%
\providecommand \Eprint [0]{\href }%
\providecommand \doibase [0]{http://dx.doi.org/}%
\providecommand \selectlanguage [0]{\@gobble}%
\providecommand \bibinfo  [0]{\@secondoftwo}%
\providecommand \bibfield  [0]{\@secondoftwo}%
\providecommand \translation [1]{[#1]}%
\providecommand \BibitemOpen [0]{}%
\providecommand \bibitemStop [0]{}%
\providecommand \bibitemNoStop [0]{.\EOS\space}%
\providecommand \EOS [0]{\spacefactor3000\relax}%
\providecommand \BibitemShut  [1]{\csname bibitem#1\endcsname}%
\let\auto@bib@innerbib\@empty
%</preamble>
\bibitem [{\citenamefont {Fukui}\ \emph {et~al.}(2005)\citenamefont {Fukui},
  \citenamefont {Hatsugai},\ and\ \citenamefont {Suzuki}}]{SM_suzuki_chern_2005}%
  \BibitemOpen
  \bibfield  {author} {\bibinfo {author} {\bibfnamefont {T.}~\bibnamefont
  {Fukui}}, \bibinfo {author} {\bibfnamefont {Y.}~\bibnamefont {Hatsugai}}, \
  and\ \bibinfo {author} {\bibfnamefont {H.}~\bibnamefont {Suzuki}},\ }\href
  {\doibase 10.1143/JPSJ.74.1674} {\bibfield  {journal} {\bibinfo  {journal}
  {J.\ Phys.\ Soc.\ Jpn.}\ }\textbf {\bibinfo {volume} {74}},\ \bibinfo {pages}
  {1674} (\bibinfo {year} {2005})}\BibitemShut {NoStop}%
\bibitem [{\citenamefont {Legner}\ \emph {et~al.}(2014)\citenamefont {Legner},
  \citenamefont {R\"uegg},\ and\ \citenamefont
  {Sigrist}}]{SM_legner_topological_2014}%
  \BibitemOpen
  \bibfield  {author} {\bibinfo {author} {\bibfnamefont {M.}~\bibnamefont
  {Legner}}, \bibinfo {author} {\bibfnamefont {A.}~\bibnamefont {R\"uegg}}, \
  and\ \bibinfo {author} {\bibfnamefont {M.}~\bibnamefont {Sigrist}},\ }\href
  {\doibase 10.1103/PhysRevB.89.085110} {\bibfield  {journal} {\bibinfo
  {journal} {Phys. Rev. B}\ }\textbf {\bibinfo {volume} {89}},\ \bibinfo
  {pages} {085110} (\bibinfo {year} {2014})}\BibitemShut {NoStop}%
\bibitem [{\citenamefont {{Ye}}\ \emph {et~al.}(2013)\citenamefont {{Ye}},
  \citenamefont {{Allen}},\ and\ \citenamefont {{Sun}}}]{SM_ye_tci_2013}%
  \BibitemOpen
  \bibfield  {author} {\bibinfo {author} {\bibfnamefont {M.}~\bibnamefont
  {{Ye}}}, \bibinfo {author} {\bibfnamefont {J.~W.}\ \bibnamefont {{Allen}}}, \
  and\ \bibinfo {author} {\bibfnamefont {K.}~\bibnamefont {{Sun}}},\
  }\href@noop {} {\bibfield  {journal} {\bibinfo  {journal} {ArXiv e-prints}\ }
  (\bibinfo {year} {2013})},\ \Eprint {http://arxiv.org/abs/1307.7191}
  {arXiv:1307.7191 [cond-mat.str-el]} \BibitemShut {NoStop}%
\bibitem [{\citenamefont {Takimoto}(2011)}]{SM_takimoto_smb6_2011}%
  \BibitemOpen
  \bibfield  {author} {\bibinfo {author} {\bibfnamefont {T.}~\bibnamefont
  {Takimoto}},\ }\href {\doibase 10.1143/JPSJ.80.123710} {\bibfield  {journal}
  {\bibinfo  {journal} {J.\ Phys.\ Soc.\ Jpn.}\ }\textbf {\bibinfo {volume}
  {80}},\ \bibinfo {pages} {123710} (\bibinfo {year} {2011})}\BibitemShut
  {NoStop}%
\bibitem [{\citenamefont {{Yu}}\ \emph {et~al.}(2015)\citenamefont {{Yu}},
  \citenamefont {{Weng}}, \citenamefont {{Hu}}, \citenamefont {{Fang}},\ and\
  \citenamefont {{Dai}}}]{SM_yu_model_2014}%
  \BibitemOpen
  \bibfield  {author} {\bibinfo {author} {\bibfnamefont {R.}~\bibnamefont
  {{Yu}}}, \bibinfo {author} {\bibfnamefont {H.}~\bibnamefont {{Weng}}},
  \bibinfo {author} {\bibfnamefont {X.}~\bibnamefont {{Hu}}}, \bibinfo {author}
  {\bibfnamefont {Z.}~\bibnamefont {{Fang}}}, \ and\ \bibinfo {author}
  {\bibfnamefont {X.}~\bibnamefont {{Dai}}},\ }\href {\doibase
  10.1088/1367-2630/17/2/023012} {\bibfield  {journal} {\bibinfo  {journal}
  {New J.\ Phys.}\ }\textbf {\bibinfo {volume} {17}},\ \bibinfo {eid} {023012}
  (\bibinfo {year} {2015})}\BibitemShut {NoStop}%
\bibitem [{\citenamefont {Kim}\ \emph {et~al.}(2014)\citenamefont {Kim},
  \citenamefont {Kim}, \citenamefont {Kang}, \citenamefont {Kim}, \citenamefont
  {Choi}, \citenamefont {Kang}, \citenamefont {Denlinger},\ and\ \citenamefont
  {Min}}]{SM_kim_termination_2014}%
  \BibitemOpen
  \bibfield  {author} {\bibinfo {author} {\bibfnamefont {J.}~\bibnamefont
  {Kim}}, \bibinfo {author} {\bibfnamefont {K.}~\bibnamefont {Kim}}, \bibinfo
  {author} {\bibfnamefont {C.-J.}\ \bibnamefont {Kang}}, \bibinfo {author}
  {\bibfnamefont {S.}~\bibnamefont {Kim}}, \bibinfo {author} {\bibfnamefont
  {H.~C.}\ \bibnamefont {Choi}}, \bibinfo {author} {\bibfnamefont {J.-S.}\
  \bibnamefont {Kang}}, \bibinfo {author} {\bibfnamefont {J.~D.}\ \bibnamefont
  {Denlinger}}, \ and\ \bibinfo {author} {\bibfnamefont {B.~I.}\ \bibnamefont
  {Min}},\ }\href {\doibase 10.1103/PhysRevB.90.075131} {\bibfield  {journal}
  {\bibinfo  {journal} {Phys. Rev. B}\ }\textbf {\bibinfo {volume} {90}},\
  \bibinfo {pages} {075131} (\bibinfo {year} {2014})}\BibitemShut {NoStop}%
\bibitem [{\citenamefont {Baruselli}\ and\ \citenamefont
  {Vojta}(2014)}]{SM_baruselli_scanning_2014}%
  \BibitemOpen
  \bibfield  {author} {\bibinfo {author} {\bibfnamefont {P.~P.}\ \bibnamefont
  {Baruselli}}\ and\ \bibinfo {author} {\bibfnamefont {M.}~\bibnamefont
  {Vojta}},\ }\href {\doibase 10.1103/PhysRevB.90.201106} {\bibfield  {journal}
  {\bibinfo  {journal} {Phys. Rev. B}\ }\textbf {\bibinfo {volume} {90}},\
  \bibinfo {pages} {201106} (\bibinfo {year} {2014})}\BibitemShut {NoStop}%
\end{thebibliography}
\end{document}